\documentclass{turabian-thesis}
\usepackage{graphicx}
\graphicspath{ {./fig/} }
\usepackage{endnotes}
\usepackage{float}

\usepackage[utf8]{inputenc}
\usepackage{csquotes, ellipsis}

\usepackage[pass, letterpaper]{geometry}

\usepackage{biblatex-chicago}
\title{Learning to Drive on the Wrong Side of the Road}
\subtitle{How American Computing Came to Rely on Conferences for Primary Publication}
\author{Elijah Bouma-Sims}
\date{Spring, 2021}

\begin{document}

\frontmatter
\maketitle

\tableofcontents
\listofillustrations

\mainmatter
\chapter*{Introduction}
\addcontentsline{toc}{chapter}{Introduction}

In 1999, the Computing Research Association released a two-page best practices memo titled ``Evaluating Computer Scientists and Engineers For Promotion and Tenure,'' which recommended that administrators consider both journal publications and other ``artifacts'' in judging the success of faculty in the field. It stated that ``Conference publication is both rigorous and prestigious'' and that ``Relying on journal publications as the sole demonstration of scholarly achievement… ignores significant evidence of accomplishment in computer science and engineering.'' The memo pulled its justification heavily from a 1994 report from the National Research Council titled Academic Careers for Experimental Computer Scientists and Engineers, which, in turn, used professional surveys to understand the nature of the field. Further, the memo's authors were prominent researchers Dr. David Patterson of UC Berkeley, Dr. Larry Snyder of the University of Washington, and Dr. Jeffery Ullman of Stanford University. Thus, while there was still controversy associated with the practice, the memo was not meant to change the landscape of computing research dramatically. Instead, it intended to help administrators understand it as it was—a field in which publications in reputable conference proceedings were valued as much or even more than publications in journals, and traditional performance metrics did not apply.
\footnote{\cite{1999CRAMemo}} 

\begin{figure}
\centering
\includegraphics[width=14cm]{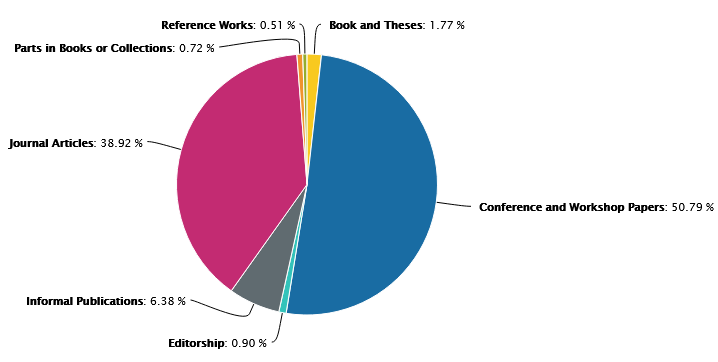}
\caption{Distribution of Publication Type in the dblp Computer Science Citation Database}
\label{dblp_graph}
\end{figure}

This statement remains true even two decades later. According to the 2019 SCImago Journal Rank (SJR), a free scientific publication ranking provided by Scimago Labs, the most impactful\footnote{The rank is determined recursively, using both the number of citations which a journal receives and the prestige of citing journals.} publication in Computer Science in 2019 was a conference proceeding: the Proceedings of the IEEE International Conference on Computer Vision.\footnote{\cite{scimagolab}} From a raw numbers perspective, 2,520,597 out of the 4,893,920 publications recorded in the dblp computer science citation database were distributed in conference proceedings.\footnote{\cite{dblp}} Figure \ref{dblp_graph} shows the distribution of publications in dblp as of October 29th, 2020.  This state of affairs stands, in contrast, to even closely related fields like electrical engineering.  In the top 50 of the 2019 SJR for electrical engineering, only four of the publications are conference proceedings, two of which are co-listed under computer science.\footnote{\cite{scimagolab}}

This state of affairs leads to a natural question: how did computing come to diverge from its predecessor fields and value conferences so highly? Naturally, much research specifically on the history of computing focuses on the impact of the development of computers, rather than discussion of the field of academic research as an independent entity. While understandable, the practical result of this focus is that little work has even analyzed the development of the academic discipline of computing, let alone the specific topic of its publication practices. For example, one of the most accessible works on the history of computing, \textit{Computer: A History of the Information Machine} by Campbell-Kelly et al., discusses the academic origins of the computer but leaves little room for the academic discipline in its later chapters.\footnote{\cite{history_of_information_machine}}

This is not to say that no literature has focused on computing as an academic discipline. The 2016 book \textit{Communities of Computing: Computer Science and Society in the ACM}  by Thomas J. Misa contains several chapters discussing the academic history of computing, including one on early female pioneers who pursued doctorates in the field.\footnote{\cite{communities_of_computing}} \textit{The Science of Computing: Shaping a Discipline} by Matt Tedre endeavors to show how researchers saw the field of computing overtime, describing three academic debates which shaped the field.\footnote{\cite{science_of_computing}} Neither book delves into the topic of publication practices, however, with each mentioning conferences more as social events than as the archival, which have become extraordinarily important in computing.

While there is a lack of literature in history relating to computing’s unique use of conferences, some discussion has occurred in other fields. Computing’s oddity in this regard makes it fertile ground for library and information science research, with researchers using bibliometric techniques to confirm computing publications' unique character. Most recently, a 2018 article titled ``Author‐based analysis of conference versus journal publication in computer science'' by Jinseok Kim confirmed that, in computing, ``conference articles seem to serve as a distinct channel of scholarly communication, not a mere preceding step to journal publications.''\footnote{\cite{jinseok2019}} A 2011 article, ``Development of computer science disciplines: a social network analysis approach'' by Pham et al. used network theory to find that ``conferences constitute social structures that shape computer science knowledge.''\footnote{\cite{pham2011}} While not historical, such research is important, as it shows that there are practical and observable differences in the way computing researchers publish when compared to other fields.

Arguably the most consequential study on computing’s use of conferences was published in 1994 and was a study commissioned by the National Research Council to address ``the challenges faced by experimental computer scientists\footnote{Experimental computer science is defined by the book as ``the building of, or the experimentation with or on, nontrivial hardware or software systems''} and engineers in academia.''\footnote{\cite{academic_carrers_for_ecsc} Pg. vii}  The report pulled on the experience of its authors and informal surveys to characterize the field. With respect to publication practices, the report stated that ``a substantial majority of respondents to the CRA-CSTB survey of [experimental computing] faculty preferred conferences as the means of dissemination by which to achieve maximum intellectual impact primarily because of timeliness.'' This observation, combined with others on different means of publication, led the report to conclude that ``a candidate's record of publication in archival journals is only one aspect of the individual's overall portfolio, and for [experimental computing] perhaps a misleading one at that.”\footnote{Ibid. Pg. 89} 5 years after publication, the report would become the basis of the aforementioned 1999 CRA memo on ``Evaluating Computer Scientists and Engineers For Promotion and Tenure.'' While certainly not as scientifically rigorous as the bibliometric studies, the publication’s significance thus shapes the subsequent discussions on computing publication.

The venue with the most discussion on computing’s publication practices is, understandably, the editorial pages of academic journals and conferences. The most recent discussion was sparked by a 2009 letter in the journal \textit{Communications of the ACM} from then editor-in-chief Moshe Y. Vardi titled ``Conferences vs. Journals in Computing Research.''\footnote{\cite{Moshe2009}} The letter spawned numerous dueling editorials discussing the topic with titles like ``Time for computer science to grow up''\footnote{\cite{Fortnow2009}} and ``Journals for certification, conferences for rapid dissemination.''\footnote{\cite{Halpern2011}} Interestingly, while none of the editorials delve into history, they all take for granted that, as Vardi puts it, ``Conference publication has had a dominant presence in computing research since the early 1980s''\footnote{\cite{Moshe2009}} and that the CRA memo on the topic essentially ended the conversation. These are taken as historical facts but not evidenced.

Perhaps the scholar who has dedicated the most energy to discussing conferences in computing is Jonathan Grudin, a pioneering computer-human interaction researcher. Grudin has written several articles on the topic and has tried to address the historical roots of conference primacy. In 2011, in the \textit{Communications of the ACM}, Grudin wrote an article titled ``Technology, Conferences, and Community,'' stating that ``the availability of text editing or word processing among computer scientists enabled the relatively inexpensive production of decent-looking proceedings before a conference'' and enabled ``Papers in ACM conferences''  to be ``widely distributed and effectively archival.''\footnote{\cite{Grudin2011}} Like others who have discussed this topic, Grudin does not provide historical evidence of his conclusions, but the technological determinism underlying the argument has a certain appeal even without investigation. Grudin’s thoughts and other editorials illustrate what forms the common “myth” of conference development in computing and serve as a reference point for analysis in my research.

Ultimately, virtually no historical research has been done discussing the unique character of Computing’s publication culture. That discussion which has occurred is unsourced and not historical research. With this in mind, this thesis attempts to trace the development of Computing’s publication culture, asking the fundamental question: how did computing develop in this unique way? Existing scholarly work does have value to this research in that it shows that there is a meaningful difference between computing and other fields, at least as far back as 1994. The editorials from computing professionals on the topic also serve to show a basic myth regarding the conference primacy in computing, which my research will serve to confirm or dispute.

Fundamentally, the historical questions at the heart of this thesis relate to how the professions in computing manage their academic knowledge. This language of ``professions'' is borrowed from Andrew Abbott’s 1988 seminal work of sociology: \textit{The System of Professions: An Essay on the Division of Expert Labor}. In the book, Abbott defines professions as ``exclusive occupational groups applying somewhat abstract knowledge to particular cases.''\footnote{\cite{abbot1988} pg. 8} Within his framework, academics are the part of the profession which ``demonstrate the rigor, the clarity, and the scientifically logical character of professional work, thereby legitimating that work in the context of larger values.''\footnote{Ibid pg. 38} While Abbott’s focus is more on how different professions gain jurisdiction within the eponymous system of professions than the publication culture of any one field, his work is useful for its bottom-up analysis of changes within professions. Professional societies like the Association of Computing Machinery (ACM) and the Computing Research Association (CRA) may seem to drive decisions in a field, but the reality is that changes in professions are driven by the complex interactions of individuals, with the more formal establishment of rules coming years after a change as been made.

Abbott’s description of the more populist reality of professional culture led to the methodology employed in this thesis. While it may be possible to attack my questions with a greater focus on archival sources, the story of computing’s publication culture is one of personal experiences which cannot be purely understood by reading ACM meeting minutes or bibliometric graphs.  In addition to traditional archival sources, I employ a series of interviews combined with more traditional sources to understand how individuals understood the changes in publication culture, creating a historical narrative with help from the patchwork of oral histories.

For this thesis, I have conducted open-ended interviews focused on the topic of publication culture with eight current professors who received their Ph.D. in computer engineering or computer science in 1980 or 1985. Interviewees were identified using the ProQuest Thesis and Dissertation database and contacted through their institutional email. Interviews were conducted over the video conferencing software, Zoom, and then transcribed from a recording. The interviews focus on how the individual perceives changes in computing’s publication culture over time and their personal preferences for publication (See the Appendix for the complete list of interview questions). While this small sample cannot possibly capture the full breadth of the field, when combined with editorials and other perspectives, they provide a unique view into the development of the field in the United States.\footnote{This protocol was approved by NC State's Institutional Review Board (IRB) as required for projects involving human subjects.}

Through this research, I have found that, while there was a rise in conference publications in the 1980s, conferences have always held an important role in computing research. Conferences emerged in an early field with few practitioners and even fewer researchers, allowing the small field to stay connected. Industrial researchers cared less about the prestige of a citation and valued presentation at a national conference over a journal credit. As the field grew and diversified, their proceedings provided page space for more and more papers as journals could not keep up. By the late 1980s, with little difference between distribution levels through reviewed conference proceedings and peer-reviewed journals, the material difference between the two types of publications was minimized. There was little reason for sub-fields that had their most prestigious publications in conferences to shift to journals. In recent times, as both conferences and journals have become widely available online, these differences have shrunk even more.

The remainder of this thesis proceeds chronologically. The first chapter discusses the early days of computing publication, starting by briefly touching on the development of modern scientific publication practices. In these early days of computing, the field was small, with access restricted if, by nothing else, a lack of computing resources and academic programs dedicated to computing. The national computing conference formed under the American Federation of Information Processing Societies is discussed as well. The second chapter shifts to discuss publications as computing diversified in the 1960s and 1970s. As the field grew, a subdivision was required to maintain community and provide space for more specific research. The third chapter discusses the remainder of the 20th century and the perspectives of my interviewees. This chapter most directly seeks to analyze the conventional narrative about conference publication's rise to prominence. Finally, the conclusion discusses the implications of this analysis and the current frontier of scientific publication.
\chapter{Early Computing Publication (1800-1970)}
\section{The Prehistory (1800-1890)}

According to Alex Csiszar in his book \textit{The Scientific Journal: Authorship and the Politics of Knowledge in the Nineteenth Century}, ``The modern scientific journal is largely an invention of the nineteenth century.''\footnote{\cite{Csiszar2018} pg. 4} While the oldest continuously operating scientific journal---widely considered to be \textit{Philosophical Transactions}---began publication in 1665,\footnote{\cite{Andrade1965}} scientific journals developed many of their common properties during this period:

\blockquote{By the early twentieth century, most scientific journals were supposed to be made up largely of papers that were original contributions to knowledge: their central claims were not to be speculative opinions nor synthetic reviews of others’ work. The latter were signed by authors who took primary credit and responsibility for their contents, but they were also expected to highlight their reliance on other authors through citations. Journals might be published either by a society or as for-profit ventures (or both), but authors did not normally receive payment for their contributions.\footnote{\cite{Csiszar2018} pg. 4} }
\nocite{kronick1962}
Before this point, research was distributed in myriad ways: lectures, books, or even private letters. The process often excluded those who were not of privileged backgrounds or failed to credit all those involved in the research. Starting as commercial ventures but eventually being adopted by even scientific elites, journals represented an ideological bend towards democracy in scientific communication after the French revolution.\footnote{Ibid.} It could be argued that conferences were the original form of scientific publication, with in-person meetings of the scientific elite being the first preferred method of sharing science. The journal, however, upended this hierarchy.

Working in this developing culture of scientific journals was arguably the first computer scientist: Charles Babbage. Babbage was interested in the mathematical tables used to speed up information processing. Inspired by human labor mechanization during the Industrial Revolution, he sought to mechanize mathematical tables' production. Babbage was tragically overambitious and unable to make his fully designed difference engine or his even more ambitious analytical engine a reality before his funding from the British government ran out. The mechanical precision required to create parts of a programmable proto-computer was unobtainable when he began his work. Babbage was left to designing machines on paper alone for the remainder of his life.\footnote{\cite{history_of_information_machine} pgs. 4-12, 41-46} 

Where Babbage's ideas did find life was through his writings. Articles on Babbage's machines appeared in journals around the world, including \textit{Philosophical Magazine}, \textit{Memoirs of the Astronomical Society of London}, and the \textit{Edinburgh Review}.\footnote{\cite{babbage_2010}} Dr. Dionysius Lardner's article in the \textit{Edinburgh Review} titled ``BABBAGE'S CALCULATING ENGINE'' resulted in the creation of a difference engine by a Swedish printer Georg Scheutz, completed in 1855.\footnote{Ibid.} This international collaboration facilitated via journal reflected the ideal of scientific publication of the time. Babbage's works on the engines would be collected and published after his death by his youngest son, Henry P. Babbage, including detailed sketches and what many consider to be the first algorithm, written by Babbage's assistant.

Early publications on mechanical computation for specific tasks would continue to appear in various forms through the 19th century and into the 20th century. When published in journals, these works were placed in scientific or mathematical periodicals rather than the more specific engineering, computing, or data processing journals that developed later. An article on one of the iterations of Sir William Thomson's\footnote{He would become Lord Kelvin} tide predictor appeared in \textit{Proceedings of the Royal Society of London}\footnote{\cite{Roberts1879}}, for example. Computing had yet to become a field of its own, with the machines of this era focused on solving tasks for specific scientific or mathematical purposes. General-purpose computing remained a task for humans with the aid of tables and mechanical calculators.

In addition to the development of the difference engine and the modern scientific journal, the 19th century also contained the birth of the field most often directly linked to computing: electrical engineering. Emerging initially as a subfield of physics, its American educational infrastructure developed in the last decades of the century as the United States electrified and more diverse uses were found for electricity. The first degree in electrical engineering would be offered by the Massachusetts Institute of Technology in 1882.\footnote{\cite{1454602}} Two years later, the first volume of \textit{Transactions of the American Institute of Electrical Engineers} (AIEE) was published.\footnote{\cite{aiee1884}} While the journal was preceded by the British \textit{Journal of the Society of Telegraph Engineers} in 1872, the AIEE journal was the first American academic journal dedicated solely to electrical engineering. The AIEE journal would endure until the 1950s, when it split into multiple transactions. The AIEE would eventually merge with the Institute of Radio Engineers (IRE), which published its own electrical engineering-focused journals from 1912 until the merger in 1963.\footnote{\cite{history_of_ieee}}

Neither publication could be considered to be about ``computing'', but their pages were host to articles discussing the components that powered the electro-mechanical tabulators and, eventually, the first generation of electronic computers. The AIEE also established one of the first computing-focused societies: the "Large Scale Computing Subcommittee of the AIEE."\footnote{\cite{Lee1996}} Finally, they also demonstrate an important point: Computing's use of conference proceedings for archival purposes did not come from electrical engineering. The electrical engineering articles which laid the foundation for modern solid-state computing were in journals.

\section{First Computing Journals (1890-1968)}

As both mechanical devices and electronics advanced, tabulating and calculating machines became more complex. By 1890, the United States Government could use an electric punch card tabulating machine to complete the census rapidly.\footnote{\cite{history_of_information_machine} pg. 13-16} These and other devices may have been much faster, but they were still incapable of completing entire tasks by themselves, however. Just as in the days of Babbage, any advanced computational tasks like weather prediction had to be done by teams of human ``computers'' with the aid of machines and mathematical tables.\footnote{Ibid. pg. 50-53} It is no surprise, then, that the first journal dedicated to discussing computing focused on the ``aids to computation'' rather than the electronic computers that would come to dominate the field.

First published in 1943 by the National Research Council, the journal \textit{Mathematical Tables and Other Aids to Computation} (MTAC) set its aim in its introductory issue to ``serve as a clearing house for mathematical tables and other aids to computation'' which had ``vastly multiplied'' in the years before its publication.\footnote{\cite{rca1943}} The quarterly journal intended to serve other academic fields by providing a place for survey articles about different types of tables, reviews of specific tables or computational tools, case studies describing methods of solving mathematical problems, and responses for general questions about computational tools. The journal, at least initially, was less of a research publication than it was a catalog mixed with a review journal. 

Despite these inauspicious beginnings, the MTAC would eventually include an article describing the operation of the first electronic programmable computer: The Electronic Numerical Integrator and Computer (ENIAC). This milestone article proceeded similar publications in AIEE and IRE journals.\footnote{\cite{Goldstine1946} \\ \cite{Burks1947} \\ \cite{Brainerd1948}} Many more early electronic computers, developed with wartime energy, were described in the pages of MTAC, including the German Z3/Zeus computer,\footnote{\cite{Lyndon1947}} the California Institute of Technology electric analog computer,\footnote{\cite{McCann1949}} and others. While these computers were discussed elsewhere, MTAC was in the unique position of being the only journal dedicated to computing for nearly a decade. Eventually, as the climate of computing changed, MTAC was transferred to the American Mathematical Society and renamed to \textit{Mathematics of Computation} in 1963.\footnote{\cite{Polachek1995}}

As computing grew, the academic community surrounding it became sufficiently large enough to support journals focused on computing as a research topic on its own rather than as an accessory to other fields. First, the IRE launched a journal called \textit{Transactions of the IRE Professional Group on Electronic Computing} at the 1952 Western Electronic Show and Convention in Long Beach, California, intending to form a ``major publication in the field of digital and analog computer-engineering.''\footnote{\cite{TransactionsIRE1952}} The Association for Computing Machinery (ACM) followed soon after with an eponymous journal, the \textit{Journal of the Association of Computing Machinery} in 1954,\footnote{\cite{Williams1954}} and later a more fast-paced, technical journal titled \textit{Communications of the Association of Computing Machinery} in 1958.\footnote{\cite{Perlis1958}} They would also publish a journal of reviews of computing literature titled \textit{Computing Reviews}. Abroad, in the United Kingdom, \textit{The Computer Journal} was founded by the British Computer Society.\footnote{\cite{Mutch1958}} By 1968, according to a list published by the National Computing Centre Limited in Manchester, England, there were 480 periodicals on computing.\footnote{\cite{Weiss1972}}

\section{Early Computing Conferences (1947-1966)}

Even as computing journals grew, conferences still held an important place. In an early field where knowledgeable researchers were rare and actual computers were even rarer, meeting in person gave computing professionals a chance to exchange ideas, establish organizations, and see novel electronic devices. In 1947, for example, a series of meetings at the Massachusetts Institute of Technology and Columbia University, NY, is credited with inspiring the precursor to the ACM---the Eastern Association for Computing Machinery.\footnote{\cite{Alt1962}} Undoubtedly, many such informal meetings influenced the early days of computing. 

Beyond informal meetings, as professional societies were established and began distributing journals, they also hosted conferences. One of the first dedicated to computing was the 1951 \textit{Joint AIEE-IRE Computer Conference: Review of Electronic Digital Computers} which was planned by computing committees in each of the organizations as well as with representatives of the ACM. The conference's goal was not to invite the presentation of new research but rather for participants to gain ``useful engineering information... from the experience of the designers and users of [electronic computers].'' The papers are relatively short---around five pages each---and include a transcript of the conference discussion at the end. They range from describing the operation of different computers, like "The National Bureau of Standards eastern automatic computer," to speculating on the future of electronic computers in general. Whereas modern conferences often focus on university research, presenters came evenly from universities and industrial or government labs, reflecting that the academic field was still in its infancy.\footnote{\cite{AIEEIRE51}} 

The proceedings were explicitly intended to be archival, with its introduction going on to state that ``published account of these machines, assembled in a report of this meeting, would be of permanent value in the development of engineering knowledge of this new field of activity.''\footnote{Ibid.} Accordingly, the proceedings are typeset in a modern-looking three-column format, including black and white photos. Papers were submitted in advance---a tradition which would carry on to the future joint conferences---allowing the proceedings to be completed before the conference.\footnote{\cite{Armer1986}} Rather than just being a cheaply printed pamphlet for attendees, some serious effort went into making it look good. How wide an audience the proceedings found is unclear, but they were undoubtedly a step above those for a more informal "meeting."

Similar joint conferences would follow, usually semiannually, from 1952 to 1987. Sponsored by the AIEE, IRE, and the ACM, the conferences were first organized under the banner of the ``National Joint Computer Committee" (NJCC) and eventually as the ``American Federation of Information Processing Societies'' (AFIPS), starting in 1961.\footnote{\cite{Ware1986}} With its incorporation under AFIPS came a clear sign of more systematic review: a complete list of abstract reviewers.\footnote{\cite{afips61}} By at latest\footnote{At the latest is used here to indicate that it is possible that full paper review began before this date. I was unable to find other calls for paper for the \textit{Joint Computing Conferences} between 1961 and 1966.} in fall 1966---based on a preserved ``Call for Papers'' from the journal \textit{Simulation}---the \textit{Joint Computing Conferences} would progress from reviewing only abstracts to full paper review, requiring prospective authors to submit ``Five draft copies of the entire paper'' and accepted authors to submit ``Final versions of the papers chosen for presentation'' 2 months before the conference.\footnote{\cite{doi:10.1177/003754976600600411}} Additionally, they made back copies of their previous conferences fully available to individuals and libraries, with the back matter of proceedings including instructions for the purchase of past conference proceedings by 1963. Already by 1966, then, the largest computing conference had adopted reviewing policies that mirrored those of modern conferences and had made the research available in a permanent form via purchasable proceedings.

Even more so than the first \textit{Joint AIEE-IRE Computer Conference}, corporations dominated the proceedings of the early AFIPS run conferences. At the 1961 Eastern Joint Computing Conference, almost all authors were associated with companies like IBM or the RAND Corporation. Likewise, the papers tended to focus on applied topics like ``Information handling in the defense communications control complex'' or ``The Saturn automatic checkout system'' rather than more abstract, theoretical topics.\footnote{Ibid.} The authorship of the 1966 conference, which reviewed full papers, mirrors this pattern. While this can be seen as reflective of a lack of university programs in computing, this may also signal that corporations were happier to share research in less traditional ways than those at universities. Unbound by academic norms, industrial researchers may have ensured the existence of conferences as final venues for publication even before more university researchers entered the space. 

In order to more clearly illustrate the impact which these conference papers could have, it is helpful to look at a case study of one particularly impactful paper from the \textit{1966 Fall Joint Computing Conference}: ``Fast Fourier Transforms: for fun and profit,'' by W. Morven Gentleman of Bell Telephone Laboratory and G. Sande of Princeton University.\footnote{\cite{Gentleman1966}} The 15-page paper, which surveys the method and application of the then-novel fast Fourier transform, has been cited by scholarly sources at least 112 times and was never republished in a journal.\footnote{As of March 2021. The exact citation count is difficult to ascertain due to variation in different services methods of searching for citations. The ACM lists the citation count at 112, while Google Scholar includes many more publications, including some non-scholarly ones, and reaches 753. The actual value is somewhere in the middle. I do not list the Google Scholar results as they can be prone to duplication or manipulation (See \cite{https://doi.org/10.1002/asi.23056})} In the three years following its publication, the paper was cited by at least 26 papers in IEEE or ACM journals, including several papers in the flagship journals the \textit{Journal of the ACM}, \textit{Communications of the ACM}, and \textit{Transactions of the IEEE}. While this pales in comparison to the at least 1,120 citations of the most cited paper from October 1966 issue of the \textit{Journal of the ACM}\footnote{\cite{Rosenfeld1966}}, it reflects that papers at conferences of the 1960s could have an immediate impact without being republished. As AFIPS cofounder and UCLA Professor Robert W. Rector stated in a 1986 retrospective, "Publication of a paper in the proceedings of a [\textit{Joint Computing Conference}] was a prestigious event... the [\textit{Joint Computing Conference}] helped shape the development of the emerging computer industry."\footnote{\cite{Rector1986}} While still not the typical path for all research, the \textit{Joint Computing Conference} was already a venue of final publication for much impactful, applied research.

In addition to the joint computing conference, the ACM also hosted an annual membership conference where technical papers were presented. Initially, these papers were not formally published, with only abstracts distributed to attendees. According to ACM cofounder Franz L. Alt, "it seemed to many that the journal Mathematical Tables and Other Aids to Computation was adequately filling the need for publications in the computer area."\footnote{\cite{Alt1962} pg. 302} The drawbacks of this system were many, however, with publications unavailable to libraries and abstracts unedited. Starting in 1952 and continuing until 1987, with some breaks, these papers were packaged into proceedings in order to "make [up-to-date information] information available to a wider public in a more permanent form."\footnote{\cite{ACM52}} By 1964, just as the \textit{Joint Computing Conference} would do, the national ACM conference began to subject papers to referring for the first time, reviewing and rejecting full papers---rather than just abstracts---before presentation, instituting one of the hallmarks of full archival publication.\footnote{\cite{Forsythe1964}} 

Evidence suggests that these proceedings were not journal quality. The articles vary from short working papers to longer, more complete ones, but few received a significant number of citations. Looking at the proceedings of the 1966 conference, compared to the AFIPS \textit{Fall Joint Computing Conference} of the same year, the papers' topics and authorship seem very similar. Corporations and other non-university-affiliated labs provide much of the research. The ACM conference is more software-focused and includes slightly more "theoretical" papers but, like the AFIPS conference, seems to focus on applied computing. Perhaps in part due to the smaller membership of the ACM vs. the combined membership of AFIPS, the proceedings \textit{ACM National Conference} seem to have been of a lower quality than the \textit{Joint Computing Conferences}.  A 1968 letter to the editor corroborates this observation, suggesting broad changes to improve them and stating bluntly: "The technical papers at the ACM are significantly poorer than those" at the joint computing conferences.\footnote{\cite{Parnas1968}}

The differences between the two national conferences may also reflect an increased sensitivity to academic concerns in the ACM as compared to the IEEE or IRE. In a set of 1986 reflections on the founding of AFIPS, former Chair of the AIEE Committee on Large-Scale Computing Devices (1957-1959) and University of Pennsylvania professor Morris Rubinoff states that, at the founding of AFIPS, the "ACM was relatively new and had a completely different outlook; they were largely academic in nature, and that was their only turf." Further, the ACM pushed for language forbidding AFIPS from entering into "direct competition with activities of its member societies."\footnote{\cite{Armer1986}} The ACM's constituency of more traditional academics as compared to the broader AFIPS community, which included many more practitioners, may have ensured that their national meeting remained closer to a traditional academic conference than the \textit{Joint Computing Conference}.

\begin{figure}
\includegraphics[width=16cm]{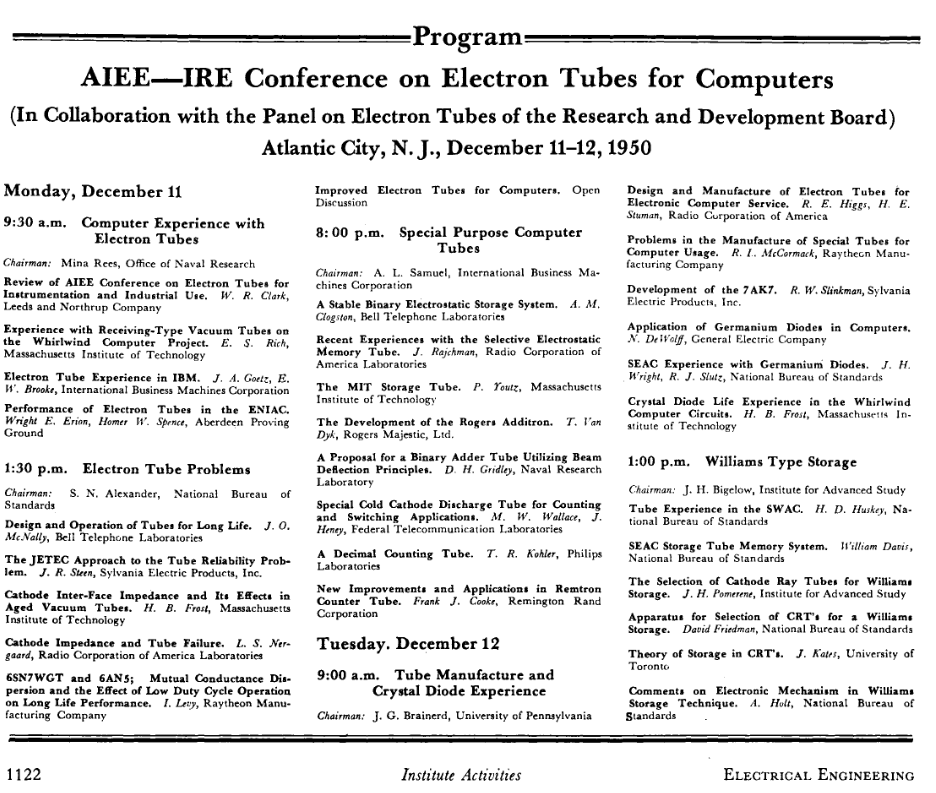}
\caption{Schedule of the 1950 AIEE-IRE Conference on Electron Tubes for Computers}
\label{aiee_ire_electron}
\end{figure}

There were also, of course, other conferences held without proceedings for which we have little record. An \textit{AIEE-IRE Conference on Electron Tubes for Computers}\footnote{\cite{AIEEIRE1950}} was held in Atlantic City, New Jersey in December 1950. There were no proceedings published from this conference. While some of the papers, such as ``The JETEC Approach to the Tube Reliability Problem'' were published later in archival journals,\footnote{\cite{Steen1951}} many were not republished. The paper ``Experience with Receiving-Type  Vacuum  Tubes on the  Whirlwind  Computer" by E.S. Rich of the Massachusetts Institute of Technology, for example, is only preserved as a report to the Office of Naval Research, set on a typewriter with attached diagrams.\footnote{\cite{rich_experience_1950}} It is difficult to fully catalog these early events without proceedings as only schedules have been preserved. Figure~\ref{aiee_ire_electron}, for example, is some of the only evidence of the \textit{AIEE-IRE Conference on Electron Tubes for Computers} besides those papers which were archived or republished. This state of affairs likely reflects the nature of the vast majority of early computing conferences---small technical meetings without archival proceedings.  
Regardless of how many small, more traditional conferences existed, it is clear that the seeds of later conference primacy were already present in these early days of computing. Conference proceedings for the most significant meetings formed simultaneously with and, in some cases, proceeded the formation of official journals. Significant research was presented at conferences, particularly from industry. Even as journals formed and grew to fill the unexplored niche, some research was still only presented at conferences and could have an immediate impact despite never being republished. 

The state of affairs with only general computing publications was not destined to last, however. As the field advanced and became more accessible, it expanded to more universities and industries, becoming more complex. At the same time, both the ACM and Institute of Electrical and Electronic Engineers (IEEE) Computer Group---which formed after the 1963 merger of the AIEE and IRE brought the AIEE Subcommittee on Large-Scale Computing Devices and the IRE Professional Group on Electronic Computers together\footnote{\cite{Letters2005}}---had more members working on diverse problems. Where once all computer research could be grouped under one umbrella, there was a need for more granularity. 
\chapter{Publications During the Diversification of Computing (1960-1980)}

\section{The ACM}

Heretofore, the professional societies of computing have remained in the discussion's background as facilitators but not the focus. However, it is worth taking a section to discuss them in-depth as they were essential to the conferences and journals that emerged to tackle specific topics. At the ACM, this took the form of ``Special Interest Groups'' (SIGs). These subdivisions provided for community as the popularity of computing exploded and, more importantly, gave space for new publications.

ACM's SIGs were established at the 15th Meeting of the ACM in 1960 when new bylaws authorized their creation as part of the larger organization.\footnote{\cite{Huskey1960}} The process for establishing one was somewhat bureaucratic. SIGs had to start life as Special Interest Committees (SICs), which would create a set of bylaws and gather members until they reached a sufficient size and funding level to become a SIG. With the formality of this process came significant benefit, however. While the ACM national organization owned their assets, the SIGs were self-governing and could manage their own finances.\footnote{\cite{Denning1971}} The first SIG was approved at most a year later, with a 1961 Letter from the ACM President declaring that the petition to accept the Special Interest Group on Mathematical Programming (SIGMAP).\footnote{\cite{Huskey1961}} By January 1962, there were at least two additional SIGS with others attempting to form.\footnote{\cite{Huskey1962}} 

The early days of SIGs were quite humble. Fitting with their intentions to connect researchers working in similar fields, they sent out plain typed newsletters to their members. The newsletters provided an overview of coming events and news relevant to the group, such as research reports or new software offerings. These publications were not archival and were explicitly not considered significant publications by the ACM Council.\footnote{\cite{Revens1972}} Indeed, the first archived SIGMAP newsletter is dated 1969 but listed as issue 5, the prior publications appearing to be lost to time.\footnote{\cite{SIGMAPBULL69}} As the SIGs grew, their newsletters would become more like quasi journals, including non-peer-reviewed, preprint/in-progress research papers.

These niche publications were significant for growth areas that had little historical precedent. Those interested in mathematical programming had journals in which they could publish. However, for the new field of computer science education, informal publication was the only dedicated page space for research and discussion. Starting with its second issue, the SIG on Computer Science Education's (SIGCSE) newsletter began to include papers describing computer science courses and teaching methods.\footnote{\cite{SIGCSEBULL69}} Their fourth issue included eight different articles and artifacts, including exams used at Carnegie Mellon.\footnote{\cite{SIGCSEBULL694}} For professors teaching the first classes of computer science students at universities and colleges across the country, these resources were undoubtedly helpful.

The fourth newsletter also included a call for papers for a \textit{Technical Symposium on Academic Education in Computer Science} sponsored by SIGCSE.\footnote{Ibid.} In addition to informal newsletters, conference organization was another role that SIGs rapidly adopted. It was an obvious fit for the national organization, as they did not have to worry about coordinating talks and workshops for a field which they may not be familiar with, offloading the work and expense to those with the most interest and experience. Appearing as a special issue of the SIGCSE newsletter, the proceedings were sent to every SIGCSE member and included 18 papers, on average seven pages in length.\footnote{\cite{SIGCSEBULL703}} SIGCSE's conference remains active 50 years later, with its 51st conference planned for 2021. SIGCSE's unique position as a group focused on pedagogical rather than technical research means that it is likely not representative of all SIGs. Still, its longevity reflects the power which SIGs grew to hold in publishing. 

Other conferences established by ACM SIGs during this time include the SIG on Symbolic and Algebraic Manipulation's sporadically scheduled \textit{Symposium on Symbolic and Algebraic Manipulation} starting in 1966;\footnote{\cite{SYMSAC66}} the SIG on Programming Language's (SIGPLAN) \textit{Conference on Language Definition} held in 1969;\footnote{\cite{SIGPLANNOT69}} the SIG on Computer Graphics and Interactive Technique's (SIGGRAPH) \textit{Annual Conference on Computer Graphics and Interactive Techniques} first held in 1974;\footnote{\cite{SIGGRAPH74}} and others. Workshops and other less formal academic meetings were also hosted by various SIGs, often in proximity to the \textit{Joint Computing Conferences} or \textit{Meeting of the ACM} to ease the burden of travel. SIGs also worked together, in some cases, hosting joint events. Throughout the late 1960s and into the early 1970s, when it came to new conferences and other academic events under ACM purview, the hub of activity was undoubtedly ACM SIGs. 

Unlike the larger national conferences, these early smaller conferences were likely never preferred for final work publication. The papers were typically only a few pages in length, and the procedure for distributing proceedings varied. Some, like the \textit{Conference on Language Definition}, are only recorded in the corresponding SIG newsletters with proceedings available for order. Others, like the first \textit{Symbosium on Symbolic and Algebraic Manipulation}, had all of their papers reprinted in full in a special issue of the \textit{Communications of the ACM}\footnote{\cite{10.1145/365758}} as well as providing independent proceedings. Others still, like the first \textit{Conference on Computer Graphics and Interactive Techniques}, simply published abstracts, with a few papers republished in other venues.\footnote{\cite{parke_model_1975}} Largely then, sharing of papers appearing in these smaller conferences occurred via republication in a journal or informal newsletter, purchase of proceedings by SIG members, or communication with the author via mail. Wider distribution was not yet guaranteed, nor were they comparable to journal publications. 

This lack of consistent distribution does not necessarily mean that these smaller conference papers did not have an impact. For some in a small sub-field, distribution to only SIG members could be enough to influence future work in the field. Only distributed as part of the conference proceedings from the 1980 \textit{Annual Conference on Computer graphics and interactive techniques} and the SIGGRAPH newsletter, the most cited paper from the conference has at least 155 citations.\footnote{According to the ACM Digital Library; as of March 2021.} Titled ``Movie-maps: An application of the optical videodisc to computer graphics'' and written by research scientist Andrew Lippman of the Massachusetts Institute of Technology, influenced many subsequent conference and journal papers.\footnote{\cite{10.1145/800250.807465}} For a researcher focused on impact rather than the prestige of citations, such results may provide an incentive to continue publishing in conferences, even as more traditional academic journals entered the space.

The \textit{ACM National Conference} continued to coexist with these other conferences throughout the 1960s and 1970s. Looking at those proceedings of the 1979 \textit{National Conference}, the impact of the papers does not seem to have changed significantly since 1966, with most papers still receiving few citations. The proceedings were still refereed and had more consistent distribution than those of the SIGs. The composition of the conference is significantly different. There were more panels and invited talks planned by SIGs. The conference proceedings are roughly a third of the length of the one held in 1966, reflecting less emphasis on academic papers and more on discussion and workshops.

The authorship of papers presented at the 1979 Conference also changed. Where papers once predominantly came from industry, there are many papers from academics based at Universities. This likely is due to the maturing of the academic field, with a more significant number of undergraduate and graduate programs in computing. It also may reflect that the constituency of the ACM was still decidedly academic. As president Daniel McCracken stated in a 1979 letter announcing a new section in \textit{Communications of the ACM}, "practitioners of computing, as distinguished from the researchers" are "the group least-well served by ACM."\footnote{\cite{10.1145/359046.359047}} More concretely, nine of the sixteen ACM presidents up to 1980 were university researchers before election, and only four did not hold a formal research position at all.\footnote{This assertion is somewhat subjective---for example, does textbook writing count as research?---but is based mainly on a mishmash of different biographical sources on each past president available via the ACM and other websites} The ACM was still a decidedly research-oriented organization in the late 1970s.

The national publication policy of the ACM was largely stagnant during the decade of the 1970s. The \textit{Journal of the ACM} (JACM) and \textit{Communications of the ACM} (CACM) were the flagship publications, with JACM appearing quarterly and CACM appearing monthly. Their respective time frames were meant to reflect the timeliness of the publications: CACM for rapid dissemination and JACM for more completed research.\footnote{\cite{Bauer1959}} In 1969, ACM's \textit{Computing Surveys} was added to provide space for computing survey articles, with original research still limited to JACM and CACM.\footnote{\cite{Aaron1969}} At this point, both JACM and CACM had become prominent publications, with JACM becoming the field's premier repository for articles on the mathematical theory. The con of this prominence was that an increasingly diverse portfolio of papers were vying for space in primarily one publication: CACM. The ACM, thus, looked to make more changes to its archival publication.

In 1969, the ACM requested comment on how it should approach its publication crunch in an editorial in CACM, although it took time for change to be evident.\footnote{\cite{Gotlieb1969}} First, in 1971, the ACM would establish a publication's board with six members to "exercise the responsibilities normally assigned to a publisher."\footnote{\cite{Weiss1971}} Then, in 1975, the first of the \textit{ACM Transactions On...} series of journals would appear, starting with \textit{ACM Transactions on Mathematical Software} which set out to "publish important results concerning mathematical software and significant computer programs"\footnote{\cite{Rice1975}} \textit{ACM Transactions on Database Systems} and \textit{ACM Transactions on Programming Languages and Systems} followed it in 1976 and 1979 respectively. More transactions would follow in the 1980s, but these were the society's initial offerings.

In this situation, we also see one of the less talked about advantages of SIG conferences. While large organizations like the ACM may be slow to act on problems---with the specialized \textit{Transactions} taking several years to materialize, even after problems were recognized---SIGs, due to their much smaller constituencies, were able to be more agile and make decisions more decisively. The ACM still bundled transactions at this point, meaning that all the organization's journals were sent to all members. Any decision to create a new journal would affect all members, not just those interested in the topic. A SIG may be established in a year and already have a conference the next, while the more extensive national ACM apparatus needs to take longer to make such decisions. By June 1979, there were 31 active SIGs with publications available to both ACM and non-ACM members.~\footnote{\cite{McCracken1979}} These more minor conference proceedings were able to gain momentum and prestige before the creation of more specific journals. 

Still, the perception of a need for more archival publications reflects the reality that conferences were not seen as a viable, permanent publication method for niche fields. Conference publications would have been used by necessity rather than preference. Simultaneously, the call for public suggestions to solve problems with the ACM's existing publication strategy hints at a failure in the existing, "traditional" publication method. Indeed, the proposal to establish the \textit{ACM Transactions on Database Systems} mentions 13 papers from the \textit{Proceedings of the 1974 annual ACM conference}, 12 papers from the \textit{AFIPS '74: Proceedings of the May 6-10, 1974, national computer conference and exposition}, and 27 papers from the Special Interest Group on Management of Data (SIGMOD) workshop, stating that many of them ``would have been submitted to and published in the [\textit{Communications of the ACM}] had there not been so many other opportunities for publication in 1974.''\footnote{\cite{Hsiao1980}} This quote suggests that for some authors of quality publications, the two national conferences and the SIGMOD workshop were acceptable venues instead of submission to \textit{Communications of the ACM}. While the author of the proposal may be overstating his case, the implication is that conferences or more specific workshops were already used by choice rather than necessity. 

A 2010 interview from Mark Mandelbaum---who worked first as Executive Editor and then Director of Publications at the ACM from 1977 to 2008---confirms the strain that existed on journals.\footnote{Interesting note: In this interview, Mandelbaum suggests that the \textit{Transactions on Graphics} and \textit{Transactions on Programming Languages and Systems} were the first two transactions released. As far as I can tell, this is a failing of his memory, as the \textit{Transactions on Graphics} was not published until 1982, and \textit{ACM Transactions on Programming Languages and Systems} was published in 1979, as mentioned above. These would have been the first two transactions that Mandelbaum oversaw, which may explain his confusion. It may also indicate that the proposals for new \textit{Transactions} shifted from outside parties to internal parties, but the overall chronology is unclear from this oral history.} In reference to a time before the \textit{Transactions}, he states that, as the popularity of CACM grew, it had a queue of over a year. Ironically, the quarterly JACM, which was intended for larger projects, only had a queue of 3 months due to its theory-focused content.\footnote{\cite{Akera2010}}

In the ACM's publication changes throughout the 1960s and 1970s, we see some momentum away from the journal hegemony that computing inherited from electrical engineering and mathematics. SIGs were established to connect the new subfields in computing. These new organizations facilitated new conferences and newsletters, which, while they may have been used out of necessity rather than choice, made room for more papers. The \textit{Transactions} series did emerge to fill that void but also took a long time to be created, leaving time for these small conferences to gain greater acceptance. Along with the continued use of the major general conferences discussed in Chapter 1, a rich ecosystem of ACM computing conferences was already in place by the mid-1970s.

\section{The IEEE Computer Group (Society)}
\nocite{King2009}

The IEEE Electronic Computer Group was also active during the period. As mentioned at the end of the previous chapter, the AIEE and the IRE merged in 1963 after an agreement by their boards in 1962.\footnote{\cite{Merger1962}} For the new society, this meant deciding how to handle some 39 journals, some of which were redundant. One carried over into the organization was the bimonthly journal \textit{IRE Transactions on Electronic Computers}, which had its moniker changed to \textit{IEEE Transactions on Electronic Computers}.  In 1966, the group released a new journal titled first \textit{Computer Group News} and then, eventually, simply \textit{Computer} which was sent to all members of the society. In 1968, the IEEE Computer Group would drop the "Electronic" from their name and the journal's name. At the same time, they also doubled the number of transactions they released in a year, moving from a bi-monthly schedule to a once-a-month schedule.\footnote{\cite{Huskey1968}} With the growth of computing, the IEEE probably felt much of the same publication crunch which the ACM did, necessitating the increase in both frequency and number of journals.

In addition to its new journal, the IEEE Computer group tried in earnest for the first time to run computer conferences in addition to the \textit{AFIPS National Computer Conferences} which they helped to run with the ACM. Christened the \textit{Annual Computer Conference} in 1967, the proceedings are not archived on IEEE XPlore, and the conference appears to have been more similar to a workshop with a narrow focus on LSI and MOS technology.\footnote{\cite{ComputerTheFirst25}}  Similarly, casual conferences would be held in 1968 and 1970, focused on similar, hardware-centric topics. These proceedings were not archived and seem to have a character more akin to those of the various SIG conferences than the \textit{ACM National Conference}.

In 1971 the IEEE Computer Group would adopt its current name and structure, becoming the IEEE Computer Society.\footnote{\cite{BirthOfAConcept}} Under IEEE governance, societies were essentially bigger, more powerful, and more general versions of the older groups. Significantly, one could join the Society without becoming a member of the IEEE, potentially bringing in a new range of people who considered themselves "scientists" rather than "engineers."\footnote{\cite{ComputerPetition}} The new society also formalized its annual computer conference, renaming it to \textit{COMPCON}.

More akin to traditional conferences, refereeing papers for the COMPCON conference of the 1970s was limited to 1000 word, "informal digests" of papers.\footnote{\cite{1978Compcon}} Each conference had a theme, such as, for the 1978 edition, "Technology of the 1980s"\footnote{Ibid.}. Reflecting the membership of the IEEE, COMPCON seems to focus more on the concerns of industry and practitioners than researchers. Only some of the full proceedings are archived, with digests of papers, lectures, and tutorials serving as the record of most COMPCONs. Ultimately, the conference cannot be considered archival, and even more so than the \textit{ACM National Conference}, seems not to have replaced the role of journals for any significant work.

The IEEE Computer Society would also dive into more formal conferences focused on specific sub-fields, including the \textit{IEEE Computer Society Computers, Software, and Applications Conference} (COMPSAC) in 1977 and the \textit{International Conference on  Distributed Computing Systems} in 1979. They also worked in partnership with ACM SIGs to run conferences. Starting in 1973 was the \textit{International Symposium on Computer Architecture} (ISCA) with the ACM SIG on Computer Architecture and, starting in 1975, was the \textit{International Conference on Software Engineering} (ICSE) with the ACM SIG on Software Engineering.\footnote{\cite{King2009}} The cross-society work with the ACM is reflective of the intersection between the IEEE Computer Society's membership with the ACM as well as the relatively larger size of the ACM compared to the IEEE Computer Society: 35,000 ACM members (1977) vs. 25,000 IEEE Computer Society members (1976).\footnote{\cite{Thompson1977} \\ \cite{ComputerTheFirst25}} While I have spoken of them in isolation of one another in this section, the two societies certainly worked together and deferred to each other in different areas.

Like the papers from the ACM SIG conferences, the IEEE conferences' papers appear to be shorter than a typical journal entry, around 5-10 pages. Additionally, in some cases, like \textit{COMPSAC}, the best papers from the conference were picked for publication in other journals.\footnote{\cite{cfp1978COMPSAC}} Unlike those from the ACM SIG conferences, there appears to be less variation in the quality and nature of the proceedings. Likely due to more centralized administration from the IEEE Computer Society, the conferences' proceedings appear to have been produced in full for many events. The proceedings for the \textit{International Conference on Software Engineering}, for example, are available online, starting from the second symposium in 1976.\footnote{\cite{10.5555/800253}} The proceedings from the earliest \textit{COMPSAC} events and \textit{International Conferences on Distributed Computing Systems} are not digitized but were sold to libraries, as will be discussed later. The constant availability of these conference proceedings may have made them more attractive, then, as repositories for final work before more specific journals emerged.

The IEEE Computer Society did also delve into specialized journal publication with its specific transactions as a part of the larger \textit{IEEE Transactions...} series, which covered topics under the electrical and electronics engineering bubble. One of the new journals established in 1975, titled \textit{IEEE Transactions on Software Engineering}, focused on the new field of software engineering. The field was popularized mainly after the NATO Science Committee held conferences to establish best computer programming practices in 1968 and 1969.\footnote{\cite{FirstIEEETSE}} Here, the word conference can more actually be subbed for working group meetings than its traditional academic meaning. They were invitation-only and centered around relevant experts' opinions rather than a broad program of papers.\footnote{\cite{naur1976software}} The second new transaction from the IEEE Computer Society was the 1979 \textit{IEEE Transactions on Pattern Analysis and Machine Intelligence}. Both these journals occupied distinct niches from those offered by coexisting ACM journals, reflecting the broadening scope of computing and the necessity for more article space.

This development---along with the specialized conferences---was driven by the IEEE Computer Society's Technical Committees. Like the ACM SIGs, these technical committees provided more focused activity under the larger banner of computer society. While they had existed in different forms long before the creation of the IEEE, 11 of them were formalized with the creation of the new Computer Group under the IEEE. They ranged from the pure hardware-focused Logic and Switching Circuit Theory Technical Committee to the software focused Programming Committee.\footnote{\cite{ComputerTheFirst25}} 

By 1975 there were 17 Technical Committees, including a newly created one on Software Engineering. It does not seem that the technical committees launched newsletters, at least not in the same way as the SIGs did. Many do have newsletters today, but past archives dating to before 1980 do not exist. Their activity was, instead, summarized in the more significant IEEE Computer Society publications. Other than this minor point, their activities seem very similar to SIGs, filling a similar niche of providing more room for diversity in the growing field. 

Predominantly, the same publication strife found in the ACM's discussions was not found in the IEEE Computer Society of this era. In an oral history, H. True Seaborn, a former IEEE Computer Society publication staff member from 1973 to 1996, recalled the editorial backlog as "healthy."\footnote{\cite{Yost2014}} While the use of this term is ambiguous, Seaborn does not discuss the backlog as being a problem or mention complaints about publication time. This contrasts heavily with Mark Mandelbaum's recollections of a long backlog and time to publication at the same time.\footnote{\cite{Akera2010}} I have also encountered numerous editorials debating ACM publication policy from this era, with fewer in the IEEE space. Absence of evidence is not evidence of absence, so it is possible that the same problems existed, but it is also likely that the ACM, as a larger society at the time, felt more of the pain than the IEEE did. Further, IEEE did expand into more specific periodicals, so they did at least feel some need for more journal space. Ultimately, the IEEE Computer Society's actions during this period primarily reflect the same diversification in computing as the ACM.

\section{Sales of Conference Proceedings}

While scientific publishing is often considered a noble, non-profit art, publications' sales are essential drivers for both publication viability and prestige of a publication. The best papers in the world are meaningless if no one sees them. Further, this has been considered one of the main distinctions between conference proceedings and journals---beyond their format, of course. Journals traditionally have a wider distribution than conferences, with proceedings originally only targeted at the attendees. Indeed, the aforementioned computer-human interaction researcher Johnathan Grudin points to an increase in printing proceedings before the conference and sales of conference proceedings in the 1980s as a reason for their current preeminence.

\begin{figure}[H]
\includegraphics[width=16cm]{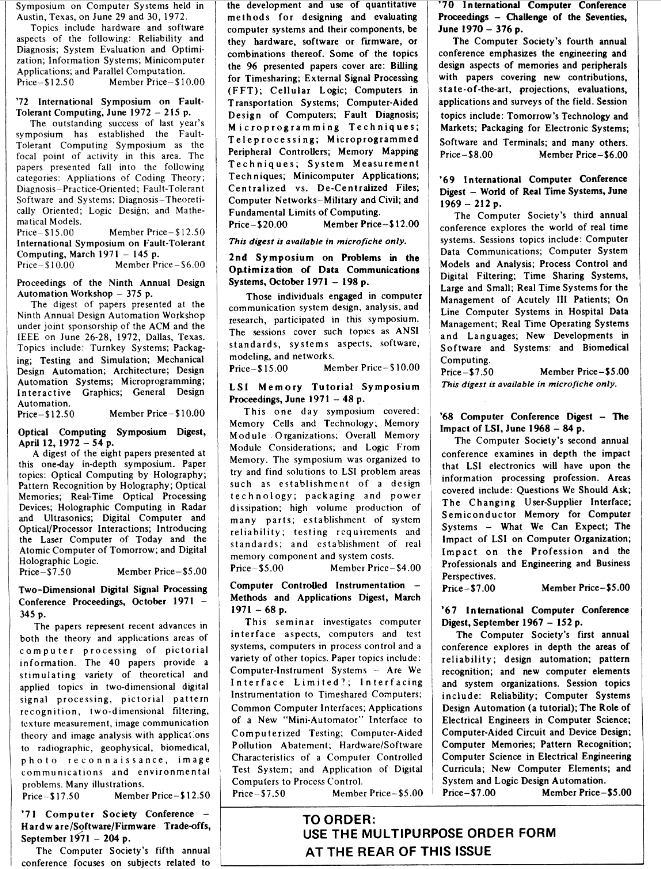}
\caption{Catalog of Past Conference Proceedings Featured in a 1973 issue of \textit{Computer}}
\label{conference_ads}
\end{figure}

The ACM and IEEE were certainly marketing and selling past conference proceedings already in the 1970s. While the distribution of proceedings could not hope to match that of a journal that is automatically sent to subscribers, aggressive marketing was pursued to sell publications, including conference proceedings. Returning to the oral history of H. True Seaborn, the publications staff member recalls the actions taken by the Computer Society to advance conference proceedings as a major source of revenue: 
\blockquote{If we had any open space in Computer we would run society house ads for current or upcoming conferences and proceedings. Or frequently conferences would schedule attendance ads or calls for papers in Computer, and we would give them a special reduced rate – lower for sponsored or co-sponsored conferences. And then periodically, Harry would send out a flyer displaying all of our past titles. That eventually became the Computer Society pubs catalog. I don’t know if we’re still issuing that or not but it became a fairly elaborate annual document of 16 or 24 pages. Harry was a super guy to work with. Throughout the 70s he and his wife Edith would routinely show up at various conferences, bringing with them a supply of conference proceedings on various related subject areas, as well as membership materials and displays. They were popular among society conference organizers and volunteers on both coasts.\footnote{\cite{Yost2014}}}
One can see a page of one of the catalogs Seaborn mentioned in figure~\ref{conference_ads}. Published in 1973, the two-page ad lists proceedings back to 1967. Interestingly, Seaborn notes that the broader IEEE was not as aggressive at selling its proceedings as the computer society was. This may hint at why conference proceedings did not grow as important in the broader electrical engineering community as they did in the computing community. 

The ACM appears to have begun to include similar catalogs within CACM in the 1970s, although it seems to be less extensive. Many just list the SIG/SIC publications without offering a method of purchase. Even once they began to include more publications, the offerings only seemed to go back a year or so. This lack of activity on the ACM's part likely hampered the proceeding's ability to be distributed widely, like journals. At the same time, it is difficult to tell what exactly was advertised at a given time as often, perhaps understandably, the sales pages or advertisements are unavailable. As discussed above, the sales of proceedings were concentrated more on a SIG level at this time, preventing their widespread prestige among the wider computing academic community. 

Regardless, the IEEE sales activity during this time helped advance the long-term distribution of conference proceedings. By selling various proceedings long past their event, conferences could be more widely read and have a greater chance of impacting future work. Libraries may also have a greater chance of purchasing them, providing them to both students and faculty. Those IEEE conferences discussed above contain several papers with dozens of citations, reflecting a greater degree of impact. From a basic economic sense, selling conference proceedings was also an excellent way to get rid of extra stock initially printed for the event. Ultimately, these good economics would advance the availability of conference proceedings, both general and niche. 

\section{Commercial Publications}

In addition to the society published publications from this era, several commercial publications emerged during this time. While generally less of a player in the conference scene, it is important to discuss them. As academic computing programs grew and the field diversified, for-profit publishers found a role, predominantly in textbooks and casual periodicals. One famous early example of a publication aimed largely at a business audience is \textit{Datamation} which ran as a technical print magazine from 1957 to 1998.\footnote{\cite{Weiss1972}} In the academic space, the venerable German publisher Springer launched a New York office in 1964,\footnote{\cite{SpringerHistory}} eventually bringing journals like \textit{Computing} (1966) and \textit{Acta Informatica} (1971) to the US Computing scene. 

These for-profit publications and others like them further helped feed a market with a growing need for page space. Still, they did not take a large share of the publishing market, with technical societies remaining at the center of the arena. Due to computing's initially narrow appeal, they were late to the computing arena and likely could not quickly gain industry market share. Additionally, it may be that academics used to an entire volunteer process throughout the editorial process were not ready for a for-profit organization to enter the middle. This is merely speculation, however. 

More prestigious were technical journals published by commercial labs like the \textit{Bell Labs Technical Journal}. Initially more of a telecommunications and electrical engineering-focused journal, the \textit{Bell Labs Technical Journal} included more computing content as the field of computers grew. Articles describing the UNIX timeshare system---a modern predecessor to many operating systems---were featured in the \textit{Bell Labs Technical Journal} starting in 1974, for example.\footnote{\cite{Ritchie1984}} These journals, while great publicity for the companies which ran them, did not provide for researchers who were not at their managing companies.

By 1980, computing was growing a rich ecosystem of journals for general topics and more specific academic sub-fields. Conferences continued to grow, covering virtually any relevant topic in the field. As academics labored to create new journals, SIG or IEEE Computing Society conferences were used either as stop gaps or, as they grew in quality, out of preference. Both SIGs and Technical Committees from the ACM and the IEEE Computer Society eventually drove forward more niche archival journals, but many established conferences endured. For-profit, academic journals emerged as well but remained in the background in the space. Finally, conference proceedings became more voluminous and archival as the IEEE heavily marketed their offerings. The stage was set for the explosion in computing facilitated through micro-computing.
\chapter{The Explosion of Computing (1980-1995)}

\section{From the Numbers}

Up to this point, this thesis has been fairly meticulous in the previous sections, carefully cataloging different conferences and journals that emerged. This process becomes nearly impossible as this thesis moves to look at the 1980s and 90s. The microprocessors developed in the 1970s fed an explosion of the personal computing market, first among hobbyists and then among the wider public. With products like the TSR-80 and the Apple II released in 1977, the late 1970s and early 1980s were the beginning of computing's transformation of modern life. Accordingly, academic programs and computing research grew.\footnote{\cite{history_of_information_machine} pg. 229-239}
 
\begin{figure}[H]
\includegraphics[width=15cm]{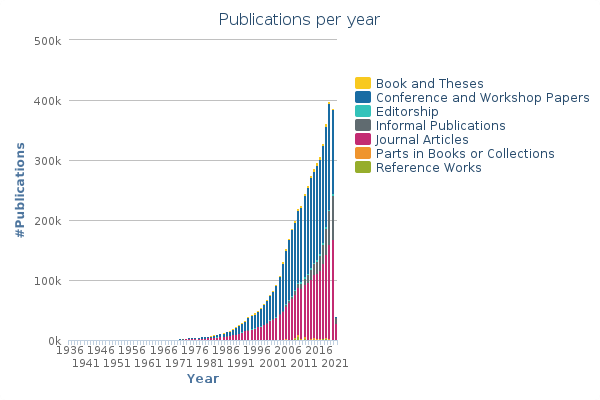}
\caption{Graph of \# of Publications by Year recorded in the dlbp database}
\label{publicationsperyear}
\end{figure}

The most acute effect of the rise of computing in academics was the staffing shortage in the field. While Ph.D. applicants and graduates remained largely stagnant, Bachelor's applicants soared, leading to a "discipline in crisis."\footnote{\cite{Denning1981}} The number of graduates would peak in 1986, as schools tightened admittance requirements as they faced untenable enrollment.\footnote{\cite{roberts2016}} The more important result for this thesis, however, is that the magnitude of research output also increased rapidly. As figure~\ref{publicationsperyear} shows, the number of publications per year in the computer science bibliography dlbp rises slowly in the 60s and 70s before exhibiting exponential growth from 1986.\footnote{\cite{dblp}. retrieved February 28th, 2020} 

To put this growth in more specific terms: before 1980, the dlbp database records only 364 conferences or workshops.\footnote{Conference and workshop are stored the same way in the dblp database; it is not practical to distinguish between the two. Workshops in computing are generally much less prestigious than conferences.} Between the years 1980 and 1990 alone, 1360 conferences or workshops are recorded in the dblp. Similarly, the dblp database contains only 45,000 journal articles published before 1980, while it contains over 80,000 journal articles published between the years 1980 and 1990 alone. In the year 1995, while the research growth had slowed, 42,906 articles were published between journals, conferences, and workshops. 

In a broad sense, the new publications that emerged centered around more granular or applied aspects of the field. This statement does not mean that all the research was wholly unprecedented but rather that the different subfields had reached sufficient critical mass to support multiple specific publications. The first issue of \textit{IEEE Transactions on Pattern Analysis and Machine Intelligence} appeared in 1979. The ACM Special Interest Group on Human-Computer Interaction (SIGCHI) was founded in 1980 and hosted the first "Conference on Easier and More Productive Use of Computer Systems" in 1981.\footnote{\cite{10.1145/800275}}. In addition to those in human-computer interaction and artificial  intelligence, more publications emerge focusing on specific subfields like graphics\footnote{\cite{4767062}}, computer networking\footnote{\cite{1702836}}, distributed systems\footnote{\cite{1667203}}, and many others. By 1995, in the dawn of the internet age, there are 432 unique conferences and workshops and 402 unique journal publications recorded in the dblp. 

Returning to the yearly publication graph from dblp, the raw numbers seem to support the idea that conferences rose to dominance as the field itself grew. Throughout the decade, the conference publication rate relative to journal publication increases, with the number of conference papers reaching near par with journal papers in 1986 and exceeding the number of journal papers in 1988. This may not give the complete picture, however. While the dblp is the best repository for meta-data on computer science papers, it is limited to the venues for which it has a ``reliable source of bibliographic meta-data.''\footnote{Ibid} Thus, many early conferences and informal publications are not recorded. \textit{The Technical Symposium on Academic Education in Computer Science} sponsored by SIGCSE described in chapter 2, for example, does not appear in dblp's database until the 17th occurrence of the conference. The lack of inclusion may indicate lesser importance, but it is not easy to know precisely how many journal or conference publications there were at a given time. To better know what publication was like for those researchers ``on the ground'' in this time, the thesis looks to the perspective of those working in the field throughout the 1980s.

\section{Perspective from Researchers}

As described in the introduction, I have conducted open-ended interviews focused on the topic of publication culture with eight current professors who received their Ph.D.  in computer engineering or computer science in 1980 or 1985. Professors who received their Ph.D. in 1980 would have been working in research from the late 1970s, while professors who received their Ph.D. in 1985 would have been working in research from the early 1980s. These individuals began their work at the turn of an era in computing and right when the conventional narrative states that conferences grew to dominance. They experienced the era of interest.

The names, institutions, and exact research areas of the individuals interviewed will remain anonymous for their privacy. For informational purposes, the traits of the group are discussed here in general terms: 
\begin{itemize}
    \item \textbf{Gender}: Two of the interviewees were female while six were male. 
    \item \textbf{Current Location}: Three of the interviewees were in the North-East; two of the interviewees were on the West Coast; two of the interviewees were in the Mid-Atlantic region; one of the interviewees was in the Mid-West. 
    \item \textbf{Ph.D. Institution}: Half received their Ph.D. at private research institutions while half received their Ph.D. at state-funded research institutions.
    \item \textbf{Current Institution}: All eight of the interviewees currently work at state-funded institutions. According to the Carnegie Classification of Institutions of Higher Education\footnote{\cite{carnegie_ranking}}, five of the interviewees' universities were classified as R1 doctoral institutions for very high research activity; one was classified as an R2 doctoral institution for high research activity; one was classified as a master's institution with small programs; and one was classified as a baccalaureate college. 
    \item \textbf{Prior Jobs}: Seven of the interviewees spent most of their time working academic jobs, while one worked predominantly non-academic jobs.
    \item \textbf{Fields of Research}: Three of the interviewees focus on educational research; two of the interviewees research artificial intelligence; two of the interviewees do theoretical research; and the other interviewee did systems research. All of these descriptions match what the interviewee currently does--not necessarily their past research.
 
\end{itemize} 

In summary, those interviewed represent a broad range of experiences, including research at both big and small schools; research in theoretical and applied computer science; and life on the East and West coasts of the United States. The only significant omission is of any professor currently at a private research institution. Some of those interviewed previously attended private research institutions, making this less of a problem. 

With few exceptions, those interviewed agreed that conferences were significant from the beginning of their research. One interviewee stated

\begin{quote}
Then as now, the emphasis was really on conference publication. There were fewer conferences, there were fewer journals. and You know, certainly. [sic] We were encouraged to. to publish in journals but we, we knew that the. [sic] The thing is that could really have a way that you could make a significant impact and the things that were really important for your career were to publish in good conferences.
\end{quote}

Alternatively, to put it much more simply, in the words of another interviewee: "Conferences were important in 1980. Conferences are important in 2020." This opinion does not match with the idea of a sudden rise of a conference in the 1980s, nor does it suggest that conference publication was a new concept. 

One of the interviewees did differ slightly. Speaking about their specific sub-field, they stated that ``[Artificial intelligence] definitely went more towards conferences and away from journals'' as the 1980s went on and that journals were perhaps worth twice a conference paper when they stated. They discuss how the most prominent artificial intelligence and machine learning overtook other publications as the decade went on. The interviewee did mention, however, that ``even when [they] started, people wanted to make an argument that for merit... conference papers [were] as significant archival and important and journals.'' This suggests that the relative stature of conferences and journals in artificial intelligence was at least in question, if not already changing, by 1980.

Another interviewee from theoretical computer science expressed a similar sentiment. They stated that "When I was first in comp sci [sic] at business meetings, there was talk of 'Oh it's bad. There are papers in conferences that never get to be in published journal form. That's a bad thing. Or that its bad that conferences count for so much' No one even talks like that anymore!" Like the previous interviewee quote, this statement suggests that there was still some degree of question around using conferences as the dominant method of dissemination in 1980. Ultimately, however, interviewees' general feel was that conferences were an important, if not a dominant, part of computing's publication culture from the earliest days.

In addition to the common sentiment about the role of conference publication from the earliest days, most interviewees shared the view that early computing researchers chose conferences for the speed of dissemination of research. However, none expressed this view emphatically, with the general sentiment that computing came to its current state through a mix of happenstance and momentum. One interviewee stated that they felt ``it was just the lack of journals and that conferences were being held. The idea of having a full paper in a conference was unusual, but that's what we had.'' That same interviewee elaborated that computing was a small field when they entered in 1975, made the use of conferences alone more practical. Another interviewee stated, ``How [computing publication] evolved overtime was not planned,'' and expressed little certainty about why conferences gained momentum. Overall, the question of how computing came to chose journals over conferences elicited more speculation than recollection from personal experience. 

None of the interviewees expressed that computing's unusual use of conferences over journals was a problem for their tenure cases. Among the several interviewees who rose into leadership roles in their respective departments, they expressed a clear sense that they needed to head off arguments about conferences before submitting a recommendation, however. One stated that he "always made sure to write a paragraph" about the difference between conferences in computing and other disciplines. Another mentioned that they had seen faculty use letters to attest to the impact of particular conference papers on the field. One interviewee also stated that he heard about the tenure cases which spawned the 1999 CRA memo, although he was not directly involved. The issue of tenure and conference publication was undoubtedly a known part of the culture of computing, even if many departments did not experience it.

The last aspect of each interview touched on the interviewee's personal preference for publication. While two of the interviewees were indifferent or expressed a preference for conferences, four out of the eight interviewees expressed the sentiment that, in an ideal world, conferences would be used for preliminary results with eventual publication in archival journals. One of the other interviewees expressed a strong preference for open access publishing, mentioning that they publish in conferences for "brownie points" with the ePrint archive, arXiv, serving as their primary mode of communication. The last interviewee spoke for a preference for publishing using books. Broadly, the sentiment seems to be that computing's publication model is at least slightly harmful. 

One of those interviewed had a radically different perspective than the others. Rather than spending their entire career in academia, they had worked as part of a research lab at a large US company for most of their life. For them, publishing papers was secondary to their primary function of assisting the internal processes. Indeed, when they went to conferences, it was more like a publicity and marketing event than a chance to share essential results. While they did work on some reasonably prominent robotics research, their ultimate career was more engineering than research. This experience matches with the analysis of the industry heavy, 1966 \textit{AFIPS} conference discussed in Chapter 1. While they were working many years after that particular event, their ambivalence towards publication quality as an industry researcher may reflect why so many were willing to publish final research papers at conferences---they did not particularly care about the perceived quality

After retiring from their corporate job, they began teaching at small colleges, mainly focusing on pedagogical research and writing reference books. When asked how publishing priorities had changed over time, they identified books as necessary for tenure cases in "many places" rather than speaking on conferences or journals. While this statement is at odds with both the other interviews and my knowledge of the field, their sentiment shows how radically one's perspective can differ within the same field. As a professor at a small, predominantly undergraduate institution, their view was much different from those of my other interviewees. 

Ultimately, the interviews conducted for this thesis slightly undercut the conventional narrative, which has conference dominance emerging in the 1980s. While those interviewed, particularly in artificial intelligence, saw a rise in the number of conferences throughout the career, they associated this more with the rise in the number of researchers than some unique explosion of conference publications. Additionally, none of the interviewees mentioned the technological advances which form the center of Grudin's narrative: word processing and easier printing. While Grudin's sense that easier printing and typesetting helped usher in more conferences may be true, those researchers interviewed here did not seem to notice, nor did they complain about conference distribution level before 1980. Indeed, only one interviewee even mentioned that conferences were not always at libraries when they began their research.

These interview findings fit with the archival record discussed in the previous two chapters, as conference proceedings from the IEEE and ACM National Conferences were already advertised and sold widely by at least by 1973, if not earlier. Smaller organizations within both societies formed conferences which continued to hold sway even after more computing journals formed. Academics who managed to write impactful work in conferences likely had little incentive to switch to newer journals. To step into one of the interviewees' shoes, it is not hard to imagine that a young researcher would enter the field in 1975 or 1980 and learn from their advisor that conferences were an acceptable or even preferable way to publish. 

The result of these interviews, then, is an unsatisfying conclusion that 'computing grew to rely on conferences because that is just what it does.' This fact, however, is a reflection of the complexity of how such norms are formed. Thousands of individual choices to publish in a conference instead of a journal eventually led to a publication culture that was ready to embrace conference primacy in the 1980s fully. By the time the interviewees entered the field, they had continued a plot that was already set in motion. 
\chapter*{Conclusion}
\addcontentsline{toc}{chapter}{Conclusion}
\setcounter{footnote}{0} 
Returning to \textit{The Scientific Journal: Authorship and the Politics of Knowledge in the Nineteenth Century} by Alex Csiszar, scientific research in general is facing a shift away from traditional publication media and journals:
\blockquote{Wherever we look, we find signs that whatever dominance the journal may have possessed is fast unraveling. Not only have authors in large segments of mathematics and physics shifted much of what they publish to the arXiv, but other fields are also experimenting with similar systems. Elsewhere, activists and editors are challenging conventions surrounding submission, prepublication review, and priority adjudication. Entrepreneurs are testing new platforms for communicating discoveries, gathering and collating feedback, and locating relevant research. Many fields are struggling with the meaning and consequences of open data policies. Since 2014, the US National Science Foundation has required principal investigators to list their ``Products'' rather than ``Publications'' on grant applications.\footnote{\cite{Andrade1965} pg. 281-282}}
The act of sending a memo urging tenure boards to consider more than just journal publications, as the CRA did in 1999, now seems somewhat quaint. That is not to say that this change does not have its own controversy associated with it, but the modern epistemological debate over the value of scholarship published in different ways is not new.
This thesis has touched on three main explanations for conference primacy in computing. The first, posited by Jonathan Grudin is one of technological dependence. In Grudin's narrative, improvements in printing technology directly led to increased printing and marketing of conference proceedings in the 1980s, particularly by the ACM. This increased availability of conference proceedings, in turn, made them more widely available and prominent, leading eventually to their present-day dominance. While appealingly direct, this explanation does not capture the whole of the evidence, and none of the interviewees discussed this point. The second explanation is pointed to by figure 3.1:  there was just too much work being produced in the 1980s; journals could not handle it, and conferences took over. Again, this explanation fails to capture some conferences' earlier prominence, nor is it entirely convincing. Many other fast-moving, young fields---such as genetics---have not shifted to a conference model to capture increased research output.

The explanation most pointed to by the evidence, then, is one of path dependence. This term comes from economics and is defined by Douglas Puffert of the University of Warwick as "the dependence of economic outcomes on the path of previous outcomes, rather than simply on current condition." In short, the explosion of computing as a research field during the 1980s combined with the existing prominence of conferences to create the modern publishing culture of American computing. Rather than consensus forming around the more traditional and accepted academic venue of journals, computing moved to rely more and more on the seemingly illogical choice of conferences, even as researchers came less from industry and more from academia. Despite this choice causing issues---such as in tenure cases---computing's history led to its modern standards.\footnote{\cite{ehencyclopedia}}

There are, of course, limitations to these conclusions. First, this study has focused entirely on the US for both scope and practical reasons. The publication culture in other countries is undoubtedly different and worthy of study in its own right. Additionally, I am limited by my access to archival material by the ongoing COVID-19 Pandemic. There is some archival material like calls for papers or older publication advertisements that may have strengthened my arguments or altered my conclusions, but I cannot access them. Finally, in my use of oral history, I am, of course, limited by those who were willing and able to speak with me, as well as by the time I had to complete interviews. While I have already captured varied opinions with only eight interviews, this thesis could only be strengthened by having more data. 

So what can practitioners learn from the story of computing's long journey to value conferences as archival? First, institutions do not change overnight. While past descriptions have found convenience in pointing to the 1980s as a single point when conferences took over, the reality is less clear. Proceedings existed and were used for permanent, final publication as early as 1952. In other places, conference proceedings were distributed as part of special editions of journals, negating the need for republication. New conferences were easy to create as nascent subfields formed in the 1960s and 70s. By the time those interviewed entered computing in the 1970s or early 1980s, conferences were already at least prominent--if not dominant.

Secondly, it is extraordinarily difficult for any individual or even collective of individuals to change a field's standards. Even as those I interviewed recognized problems in computing's publication paradigm, they seemed resigned to the status quo. Whatever becomes the future standard scientific publication---whether its traditional journals, open access publications, arXiv prints, or something I cannot yet imagine---it will likely not be dictated from standards bodies, committees of professors, or even individuals. Instead, consensus will come from a complex system of interacting stakeholders who gradually create the future of publications.

Ultimately, even in computing, the norms of publishing are continually changing. The nature of computing conferences was radically changed by the COVID-19 pandemic preventing international travel. Some in the field have proposed taking this opportunity to "Reboot the Computing-Research Publication Systems"\footnote{\cite{10.1145/3437991}} as the world returns to some form of normalcy in the latter half of 2021 and 2022. The ACM is launching a task force to explore the topic. Change, however, will depend on tens of thousands of researchers leaping to change their philosophy. 

\backmatter
\begin{appendixes}
    \chapter{Interview Questions}

Before asking questions, the participant will be informed that recording is going to begin. The semi-structured interview will center around the following questions, as well as any related follow ups:
\begin{enumerate}
    \item When did you first begin working in research?
    \item How would you describe the fields you have worked in? What has your research focused on?
    \item How has the nature of computing publication changed over time? 
    \begin{enumerate}
        \item Has the role of conferences and journal publications changed? When do you feel these changes occurred?
        \item How did this change represent itself in your work?
    \end{enumerate}
    \item With regards to the previous question, why do you think the publication culture changed?
    \item How do you prefer to have your work published? Why?
    \item Is there anything else you think I should know or include in my study?
\end{enumerate}

After the questions are completed, the participant will be informed that recording is ending. They will be reminded to contact the researchers with any questions. The interview will not last any longer than 30 minutes.

\end{appendixes}

\printbibliography

@misc{1999CRAMemo,
    Title = {Evaluating Computer Scientists and Engineers For Promotion and Tenure},
    year = {1999},
    author = {David Patterson and Lawrence Snyder and Jeffrey Ullman},
    publisher = {Computing Research Association}
    
}

@article{https://doi.org/10.1002/asi.23056,
author = {Delgado López-Cózar, Emilio and Robinson-García, Nicolás and Torres-Salinas, Daniel},
title = {The Google scholar experiment: How to index false papers and manipulate bibliometric indicators},
journal = {Journal of the Association for Information Science and Technology},
volume = {65},
number = {3},
pages = {446-454},
doi = {10.1002/asi.23056},
year = {2014}
}

@article{Rosenfeld1966,
author = {Rosenfeld, Azriel and Pfaltz, John L.},
title = {Sequential Operations in Digital Picture Processing},
year = {1966},
issue_date = {Oct. 1966},
publisher = {Association for Computing Machinery},
address = {New York, NY, USA},
volume = {13},
number = {4},
issn = {0004-5411},
doi = {10.1145/321356.321357},
journal = {J. ACM},
month = oct,
pages = {471–494},
numpages = {24}
}

@techreport{rich_experience_1950,
	type = {Technical {Report}},
	title = {Experience with {Receiving}-{Type} {Vacuum} {Tubes} on the {Whirlwind} {Computer} {Project}},
	url = {http://dome.mit.edu/xmlui/handle/1721.3/39819},
	language = {en},
	urldate = {2021-03-23},
	institution = {MIT Servomechanisms Laboratory},
	author = {Rich, Edwin S.},
	month = dec,
	year = {1950},
}

@inproceedings{Gentleman1966,
author = {Gentleman, W. M. and Sande, G.},
title = {Fast Fourier Transforms: For Fun and Profit},
year = {1966},
isbn = {9781450378932},
publisher = {Association for Computing Machinery},
address = {New York, NY, USA},
doi = {10.1145/1464291.1464352},
booktitle = {Proceedings of the November 7-10, 1966, Fall Joint Computer Conference},
pages = {563–578},
numpages = {16},
location = {San Francisco, California},
series = {AFIPS '66 (Fall)}
}

@book{babbage_2010, place={Cambridge}, edition={Digital}, title={Babbages Calculating Engines Being a Collection of Papers Relating to them ; their History and Construction}, publisher={Cambridge University Press}, author={Babbage, Charles}, editor={Babbage, Henry P.}, year={2010}}

@book{history_of_information_machine,
    Title= {Computer: A History of the Information Machine},
    Year = {2014},
    edition = {3},
    publisher = {Westview Press},
    location = {Boulder, Colorado},
    author = {Martin Campbell-Kelly and William Aspray and Nathan Ensmenger and Jeffery R. Yost}
}

@book{communities_of_computing,
    Title = {Communities of Computing: Computer Science and Society in the ACM},
    author = {Thomas J. Misa},
    publisher = {Association for Computing Machinery and Morgan \& Claypool},
    location = {New York City, New York},
    month = {11},
    year = {2016},
    doi = {10.1145/2973856},
    isbn = {978-1-970001-87-7}
}

@book{ academic_carrers_for_ecsc,
    Title = {Academic Careers for Experimental Computer Scientists and Engineers},
    author = {Committee on Academic Careers for Experimental Computer Scientists and National Research Council and Computer Science and Telecommunications Board},
    pgs = {1-151},
    publisher = {National Academies Press},
    location = {Washington, D.C.},
    year = {1994},
    day = {01},
    month = {02}
}

@book{science_of_computing,
    Title = {The Science of Computing: Shaping a Discipline},
    Author =  {Matti Tedre},
    publisher = {Chapman \& Hall/CRC},
    year = {2014},
    month = {12},
    location = {Boca Raton, Florida}

}

@ARTICLE{1702836,

  author={},

  journal={IEEE Transactions on Software Engineering}, 

  title={Call for Papers}, 

  year={1981},

  volume={SE-7},

  number={2},

  pages={257-257},

  doi={10.1109/TSE.1981.234527}}

@ARTICLE{1667203,

  author={},

  journal={Computer}, 

  title={Call for Papers}, 

  year={1981},

  volume={14},

  number={12},

  pages={54-54},

  doi={10.1109/C-M.1981.220296}}

@proceedings{10.1145/800275,
title = {CHI '81: Proceedings of the Joint Conference on Easier and More Productive Use of Computer Systems. (Part - I): Information Processing in the Social Sciences and Humanities - Volume 1981},
year = {1981},
isbn = {0897910567},
publisher = {Association for Computing Machinery},
address = {New York, NY, USA},
location = {Ann Arbor, MI}
}

@book{carnegie_ranking,
     title = {{The Carnegie Classification of Institutions of Higher Education}},
     edition = {2018 edition},
     Author = {{Indiana University Center for Postsecondary Research}},
     address = {Bloomington, IN}
}

@article{10.1145/3437991,
author = {Vardi, Moshe Y},
title = {Reboot the Computing-Research Publication Systems},
year = {2020},
issue_date = {January 2021},
publisher = {Association for Computing Machinery},
address = {New York, NY, USA},
volume = {64},
number = {1},
issn = {0001-0782},
doi = {10.1145/3437991},
journal = {Commun. ACM},
month = dec,
pages = {7},
numpages = {1}
}

@ARTICLE{4767062,

  author={},

  journal={IEEE Transactions on Pattern Analysis and Machine Intelligence}, 

  title={Call for Papers}, 

  year={1981},

  volume={PAMI-3},

  number={1},

  pages={116-116},

  doi={10.1109/TPAMI.1981.4767062}}

@online{scimagolab,
    author = {Scimago Lab},
    title = {Scimago Journal \& Country Rank},
    url  = {https://www.scimagojr.com},
    addendum = {(accessed: October 29th, 2019)}
}

@online{dblp,
    title = {dblp computer science bibliography},
    addendum = {Monthly snapshot releases of Feburary 2021 and March 2021.},
    url = {https://dblp.org/xml/release/}
}

@article{pham2011,
    author = {Manh Cuong Pham and Ralf Klamma and Matthias Jarke},
    title = {Development of computer science disciplines: a social network analysis approach},
    journal = {Soc. Netw. Anal. Min},
    volume = { 1},
    number = {4},
    pages = {321-340},
    doi = {10.1007/s13278-011-0024-x}
}

@article{jinseok2019,
author = {Kim, Jinseok},
title = {Author-based analysis of conference versus journal publication in computer science},
journal = {Journal of the Association for Information Science and Technology},
volume = {70},
number = {1},
pages = {71-82},
doi = {10.1002/asi.24079},
year = {2019}
}

@article{Moshe2009,
author = {Vardi, Moshe Y},
title = {Conferences vs. Journals in Computing Research},
year = {2009},
issue_date = {May 2009},
publisher = {Association for Computing Machinery},
address = {New York, NY, USA},
volume = {52},
number = {5},
issn = {0001-0782},
doi = {10.1145/1506409.1506410},
journal = {Commun. ACM},
month = may,
pages = {5},
numpages = {1}
}

@article{Halpern2011,
author = {Halpern, Joseph Y. and Parkes, David C.},
title = {Journals for Certification, Conferences for Rapid Dissemination},
year = {2011},
issue_date = {August 2011},
publisher = {Association for Computing Machinery},
address = {New York, NY, USA},
volume = {54},
number = {8},
issn = {0001-0782},
doi = {10.1145/1978542.1978555},
journal = {Commun. ACM},
month = aug,
pages = {36–38},
numpages = {3}
}

@article{Grudin2011,
author = {Grudin, Jonathan},
title = {Technology, Conferences, and Community},
year = {2011},
issue_date = {February 2011},
publisher = {Association for Computing Machinery},
address = {New York, NY, USA},
volume = {54},
number = {2},
issn = {0001-0782},
doi = {10.1145/1897816.1897834},
journal = {Commun. ACM},
month = feb,
pages = {41–43},
numpages = {3}
}

@article{Fortnow2009,
author = {Fortnow, Lance},
title = {Viewpoint Time for Computer Science to Grow Up},
year = {2009},
issue_date = {August 2009},
publisher = {Association for Computing Machinery},
address = {New York, NY, USA},
volume = {52},
number = {8},
issn = {0001-0782},
doi = {10.1145/1536616.1536631},
journal = {Commun. ACM},
month = aug,
pages = {33–35},
numpages = {3}
}

@book{abbot1988,
  title = {The system of professions : an essay on the division of expert labor},
  author = {Abbott, Andrew Delano},
  year = {1988},
  publisher = {University of Chicago Press},
  location = {Chicago, Illinois}
}

@article{Alt1962,
author = {Alt, Franz L.},
title = {Fifteen Years ACM},
year = {1962},
issue_date = {June 1962},
publisher = {Association for Computing Machinery},
address = {New York, NY, USA},
volume = {5},
number = {6},
issn = {0001-0782},
doi = {10.1145/367766.367777},
journal = {Commun. ACM},
month = {6},
pages = {300–307},
numpages = {8}
}

@ARTICLE{Polachek1995,  author={H. {Polachek}},  journal={IEEE Annals of the History of Computing},   title={History of the journal Mathematical Tables and other Aids to Computation, 1959-1965},   year={1995},  volume={17},  number={3},  pages={67-74},  doi={10.1109/85.397062}}

@book{Csiszar2018,
title = {The Scientific Journal: Authorship and the Politics of Knowledge in the Nineteenth Century},
author = {Alex Csiszar},
publisher = {The University of Chicago Press},
location = {Chicago, Illinois},
year = {2018},
doi = {10.7208/chicago/9780226553375.003.0001}
}

@article{Andrade1965,
 ISSN = {00359149},
 URL = {http://www.jstor.org/stable/3519880},
 author = {E. N. da C. Andrade},
 journal = {Notes and Records of the Royal Society of London},
 number = {1},
 pages = {9--27},
 publisher = {The Royal Society},
 title = {The Birth and Early Days of the Philosophical Transactions},
 volume = {20},
 year = {1965}
}

@ARTICLE{1454602,

  author={F. E. {Terman}},

  journal={Proceedings of the IEEE}, 

  title={A brief history of electrical engineering education}, 

  year={1976},

  volume={64},

  number={9},

  pages={1399-1407},

  doi={10.1109/PROC.1976.10333}}

@book{kronick1962,
 author = {Kronick, David A.},
 title = {A History of Scientific And Technical Periodicals: the Origins And Development of the Scientific And Technological Press, 1665-1790},
 location = {New York},
 publisher = {Scarecrow Press},
 year = {1962}
 }

@article{Roberts1879,
 ISSN = {03701662},
 URL = {http://www.jstor.org/stable/113758},
 author = {E. Roberts},
 journal = {Proceedings of the Royal Society of London},
 number = {},
 pages = {198--201},
 publisher = {The Royal Society},
 title = {Preliminary Note on a New Tide-Predicter},
 volume = {29},
 year = {1879}
}

@periodical{aiee1884,
title = {Transactions of the American Institute of Electrical Engineers},
issue = {1},
volume = {1},
year = {1884},
month = {5},
publisher = {American Institute of Electrical Engineers}
}

@misc{history_of_ieee,
	title = {History of {IEEE}},
	url = {https://www.ieee.org/about/ieee-history.html},
	urldate = {2020-11-20},
}

@article{rca1943,
 ISSN = {08916837},
 URL = {http://www.jstor.org/stable/2002682},
 journal = {Mathematical Tables and Other Aids to Computation},
 number = {1},
 pages = {1--2},
 publisher = {American Mathematical Society},
 title = {Introductory},
 volume = {1},
 year = {1943}
}

@article{Burks1947,

author={Burks,A. W.},

year={1947},

title={Electronic Computing Circuits of the ENIAC},

journal={Proceedings of the IRE},

volume={35},

number={8},

pages={756-767},

isbn={0096-8390},

language={English},

}

@article{Brainerd1948,

author={Brainerd,J. G. and Sharpless,T. K.},

year={1948},

title={The ENIAC},

journal={Electrical engineering (New York, N.Y.)},

volume={67},

number={2},

pages={163-172},

isbn={0095-9197},

language={English},

}

@article{Goldstine1946,
 ISSN = {08916837},
 URL = {http://www.jstor.org/stable/2002620},
 author = {H. H. Goldstine and Adele Goldstine},
 journal = {Mathematical Tables and Other Aids to Computation},
 number = {15},
 pages = {97--110},
 publisher = {American Mathematical Society},
 title = {The Electronic Numerical Integrator and Computer (ENIAC)},
 volume = {2},
 year = {1946}
}

@article{Lyndon1947,
 ISSN = {08916837},
 URL = {http://www.jstor.org/stable/2002238},
 author = {Roger C. Lyndon},
 journal = {Mathematical Tables and Other Aids to Computation},
 number = {20},
 pages = {354--359},
 publisher = {American Mathematical Society},
 title = {The Zuse Computer},
 volume = {2},
 year = {1947}
}

@article{McCann1949,
 ISSN = {08916837},
 URL = {http://www.jstor.org/stable/2002249},
 author = {G. D. McCann},
 journal = {Mathematical Tables and Other Aids to Computation},
 number = {28},
 pages = {501--513},
 publisher = {American Mathematical Society},
 title = {The California Institute of Technology Electric Analog Computer},
 volume = {3},
 year = {1949}
}

@ARTICLE{TransactionsIRE1952,

  author={},

  journal={Transactions of the I.R.E. Professional Group on Electronic Computers}, 

  title={Technical sessions on electronic computers}, 

  year={1952},

  volume={PGEC-1},

  number={1},

  pages={1-1},

  doi={10.1109/IREPGELC.1952.6499393}}

@article{Williams1954,
author = {Williams, Samuel B.},
title = {The Association for Computing Machinery},
year = {1954},
issue_date = {Jan. 1954},
publisher = {Association for Computing Machinery},
address = {New York, NY, USA},
volume = {1},
number = {1},
issn = {0004-5411},
doi = {10.1145/320764.320765},
journal = {J. ACM},
month = jan,
pages = {1–3},
numpages = {3}
}

@article{Perlis1958,
editor = {Perlis, A. J.},
year = {1958},
issue_date = {Jan. 1958},
publisher = {Association for Computing Machinery},
volume = {1},
number = {1},
issn = {0001-0782},
journal = {Commun. ACM}
}

@article{Mutch1958,
editor = {Eric N. Mutch},
year = {1958},
issue_date = {Jan. 1958},
publisher = {Oxford University Press},
volume = {1},
number = {1},
issn = {0010-4620},
journal = {The Computer Journal}
}

@article{Weiss1972,
author = {Weiss, Eric A},
title = {Publications in Computing: An Informal Review},
year = {1972},
issue_date = {July 1972},
publisher = {Association for Computing Machinery},
address = {New York, NY, USA},
volume = {15},
number = {7},
issn = {0001-0782},
doi = {10.1145/361454.361456},
journal = {Commun. ACM},
month = jul,
pages = {491–497},
numpages = {7}
}

@article{doi:10.1177/003754976600600411,
author = {Maurice I. Stein},
title ={Dear John},
journal = {SIMULATION},
volume = {6},
number = {4},
pages = {vii-vii},
year = {1966},
doi = {10.1177/003754976600600411},
}

@ARTICLE{Armer1986,

  author={P. {Armer} and M. M. {Astrahan} and I. L. {Auerbach} and W. M. {Carlson} and A. A. {Cohen} and M. R. {Fox} and C. A. R. {Kagan} and M. {Rubinoff} and J. {Sherman} and W. H. {Ware}},

  journal={Annals of the History of Computing}, 

  title={Reflections on a Quarter-Century: AFIPS Founders}, 

  year={1986},

  volume={8},

  number={3},

  pages={225-256},

  doi={10.1109/MAHC.1986.10047}}

@ARTICLE{Ware1986,

  author={W. H. {Ware}},

  journal={Annals of the History of Computing}, 

  title={AFIPS in Retrospect}, 

  year={1986},

  volume={8},

  number={3},

  pages={303-310},

  doi={10.1109/MAHC.1986.10052}}

@ARTICLE{AIEEIRE1950,

  author={},

  journal={Electrical Engineering}, 

  title={AIEE-IRE conference on electron tubes for computers}, 

  year={1950},

  volume={69},

  number={12},

  pages={1122-1122},

  doi={10.1109/EE.1950.6437165}}

@proceedings{AIEEIRE51,
title = {AIEE-IRE '51: Papers and Discussions Presented at the Dec. 10-12, 1951, Joint AIEE-IRE Computer Conference: Review of Electronic Digital Computers},
year = {1951},
isbn = {9781450378512},
publisher = {Association for Computing Machinery},
address = {New York, NY, USA},
location = {Philadelphia, Pennsylvania}
}

@proceedings{afips61,
title = {AFIPS '61 (Eastern): Proceedings of the December 12-14, 1961, Eastern Joint Computer Conference: Computers - Key to Total Systems Control},
year = {1961},
isbn = {9781450378734},
publisher = {Association for Computing Machinery},
address = {New York, NY, USA},
location = {Washington, D.C.}
}

@ARTICLE{Steen1951,

  author={J. R. {Steen}},

  journal={Proceedings of the IRE}, 

  title={The JETEC Approach to the Tube-Reliability Problem}, 

  year={1951},

  volume={39},

  number={9},

  pages={998-1000},

  doi={10.1109/JRPROC.1951.273739}}

@proceedings{ACM52,
title = {ACM '52: Proceedings of the 1952 ACM National Meeting (Pittsburgh)},
year = {1952},
isbn = {9781450373623},
publisher = {Association for Computing Machinery},
address = {New York, NY, USA},
location = {Pittsburgh, Pennsylvania}
}

@ARTICLE {Letters2005,
author = {},
journal = {Computer},
title = {Letters},
year = {2005},
volume = {24},
number = {09},
issn = {1558-0814},
pages = {6-7},
doi = {10.1109/MC.2005.305},
publisher = {IEEE Computer Society},
address = {Los Alamitos, CA, USA},
month = {sep}
}

@article{Huskey1960,
author = {Huskey, Harry D},
title = {Letter from the President of ACM},
year = {1960},
issue_date = {September 1960},
publisher = {Association for Computing Machinery},
address = {New York, NY, USA},
volume = {3},
number = {9},
issn = {0001-0782},
doi = {10.1145/367390.367395},
journal = {Commun. ACM},
month = sep,
pages = {481},
numpages = {1}
}

@article{Denning1971,
author = {Denning, Peter J},
title = {On ACM Special Interest Groups and Committees},
year = {1971},
issue_date = {Nov. 1971},
publisher = {Association for Computing Machinery},
address = {New York, NY, USA},
volume = {14},
number = {11},
issn = {0001-0782},
doi = {10.1145/362854.362873},
journal = {Commun. ACM},
month = nov,
pages = {694–696},
numpages = {3}
}

@article{Weiss1971,
author = {Weiss, Eric A},
title = {On the ACM Publications Board},
year = {1971},
issue_date = {July 1971},
publisher = {Association for Computing Machinery},
address = {New York, NY, USA},
volume = {14},
number = {7},
issn = {0001-0782},
doi = {10.1145/362619.362621},
journal = {Commun. ACM},
month = jul,
pages = {441–442},
numpages = {2}
}

@article{Revens1972,
author = {Revens, Lee},
title = {The First Twenty-Five Years: ACM 1947-1962},
year = {1972},
issue_date = {July 1972},
publisher = {Association for Computing Machinery},
address = {New York, NY, USA},
volume = {15},
number = {7},
issn = {0001-0782},
doi = {10.1145/361454.361455},
journal = {Commun. ACM},
month = jul,
pages = {485–490},
numpages = {6}
}

@article{Huskey1961,
author = {Huskey, Harry D},
title = {Letter from the President},
year = {1961},
issue_date = {Oct. 1961},
publisher = {Association for Computing Machinery},
address = {New York, NY, USA},
volume = {4},
number = {10},
issn = {0001-0782},
doi = {10.1145/366786.366787},
journal = {Commun. ACM},
month = oct,
pages = {415},
numpages = {1}
}

@article{Huskey1962,
author = {Huskey, Harry D},
title = {Letter from the President of ACM},
year = {1962},
issue_date = {Jan. 1962},
publisher = {Association for Computing Machinery},
address = {New York, NY, USA},
volume = {5},
number = {1},
issn = {0001-0782},
doi = {10.1145/366243.366267},
journal = {Commun. ACM},
month = jan,
pages = {1},
numpages = {1}
}

@periodical{SIGMAPBULL69,
year = {1969},
issue_date = {June 1969},
publisher = {Association for Computing Machinery},
number = {5},
issn = {0163-5786},
title = {SIGMAP Bull.}
}

@ARTICLE{Rector1986,

  author={R. W. {Rector}},

  journal={Annals of the History of Computing}, 

  title={Personal Recollections on the First Quarter-Century of AFIPS}, 

  year={1986},

  volume={8},

  number={3},

  pages={261-269},

  doi={10.1109/MAHC.1986.10053}}

@proceedings{10.5555/800253,
title = {ICSE '76: Proceedings of the 2nd International Conference on Software Engineering},
year = {1976},
publisher = {IEEE Computer Society Press},
address = {Washington, DC, USA},
location = {San Francisco, California, USA}
}

@inproceedings{10.1145/800250.807465,
author = {Lippman, Andrew},
title = {Movie-Maps: An Application of the Optical Videodisc to Computer Graphics},
year = {1980},
isbn = {0897910214},
publisher = {Association for Computing Machinery},
address = {New York, NY, USA},
doi = {10.1145/800250.807465},
booktitle = {Proceedings of the 7th Annual Conference on Computer Graphics and Interactive Techniques},
pages = {32–42},
numpages = {11},
location = {Seattle, Washington, USA},
series = {SIGGRAPH '80}
}

@article{cfp1978COMPSAC,
year={1978},
month={03},
title={Call for Papers: [1]},
volume={4},
number={2},
pages={148},
note={Copyright - Copyright IEEE Computer Society Mar 1978; Last updated - 2020-11-17; CODEN - IESEDJ},
isbn={00985589},
language={English},
}

@article{1978Compcon,
year={1978},
month={01},
title={Call for Papers},
journal={IEEE Transactions on Software Engineering},
volume={4},
number={1},
pages={78},
isbn={00985589},
language={English},
url={https://proxying.lib.ncsu.edu/index.php/login?url=https://www-proquest-com.prox.lib.ncsu.edu/scholarly-journals/call-papers/docview/195580905/se-2?accountid=12725}
}

@article{10.1145/359046.359047,
author = {McCracken, Daniel D},
title = {ACM President's Letter: The ACM in 1979: Upward and Outward},
year = {1979},
issue_date = {Jan. 1979},
publisher = {Association for Computing Machinery},
address = {New York, NY, USA},
volume = {22},
number = {1},
issn = {0001-0782},
doi = {10.1145/359046.359047},
journal = {Commun. ACM},
month = jan,
pages = {1–2},
numpages = {2}
}

@article{ehencyclopedia,
author = {Puffert, Douglas},
title = {Path Dependence},
editor = {Whaples, Robert},
date = {Feb. 10, 2008},
url = {http://eh.net/encyclopedia/path-dependence/},
journal = {EH.Net Encyclopedia}
}

@article{McCracken1979,
author = {McCracken, Daniel D},
title = {ACM President's Letter: “the Benefits of ACM Membership”},
year = {1979},
issue_date = {June 1979},
publisher = {Association for Computing Machinery},
address = {New York, NY, USA},
volume = {22},
number = {6},
issn = {0001-0782},
doi = {10.1145/359114.359115},
journal = {Commun. ACM},
month = jun,
pages = {333–334},
numpages = {2}
}

@article{Forsythe1964,
author = {Forsythe, George E.},
title = {President's Letter to the ACM Membership},
year = {1964},
issue_date = {Aug. 1964},
publisher = {Association for Computing Machinery},
address = {New York, NY, USA},
volume = {7},
number = {8},
issn = {0001-0782},
doi = {10.1145/355586.364833},
journal = {Commun. ACM},
month = aug,
pages = {507–509},
numpages = {3}
}

@periodical{SIGCSEBULL69,
year = {1969},
issue_date = {June 1969},
publisher = {Association for Computing Machinery},
volume = {1},
number = {2},
issn = {0097-8418},
title = {SIGCSE Bull.}
}

@periodical{SIGCSEBULL694,
year = {1969},
issue_date = {December 1969},
publisher = {Association for Computing Machinery},
volume = {1},
number = {4},
issn = {0097-8418},
title = {SIGCSE Bull.}
}

@article{10.1145/365758,
editor = {Salton, Gerard},
year = {1966},
issue_date = {Aug. 1966},
publisher = {Association for Computing Machinery},
volume = {9},
number = {8},
issn = {0001-0782},
journal = {Commun. ACM}
}

@article{parke_model_1975,
	title = {A model for human faces that allows speech synchronized animation},
	volume = {1},
	issn = {0097-8493},
	url = {https://www.sciencedirect.com/science/article/pii/0097849375900242},
	number = {1},
	journal = {Computers \& Graphics},
	author = {Parke, Frederic I.},
	year = {1975},
	pages = {3--4},
}

@periodical{SIGCSEBULL703,
year = {1970},
issue_date = {November 1970},
publisher = {Association for Computing Machinery},
volume = {2},
number = {3},
issn = {0097-8418},
title = {SIGCSE Bull.}
}

@periodical{SIGPLANNOT69,
year = {1969},
issue_date = {January 1969},
publisher = {Association for Computing Machinery},
volume = {4},
number = {1},
issn = {0362-1340},
title = {SIGPLAN Not.}
}

@proceedings{SYMSAC66,
title = {SYMSAC '66: Proceedings of the First ACM Symposium on Symbolic and Algebraic Manipulation},
year = {1966},
isbn = {9781450373715},
publisher = {Association for Computing Machinery},
address = {New York, NY, USA}
}

@proceedings{SIGGRAPH74,
title = {SIGGRAPH '74: Proceedings of the 1st Annual Conference on Computer Graphics and Interactive Techniques},
year = {1974},
isbn = {9781450373524},
publisher = {Association for Computing Machinery},
address = {New York, NY, USA},
location = {Boulder, Colorado}
}

@article{Bauer1959,
author = {Bauer, Walter F. and Juncosa, Mario L. and Perlis, Alan J.},
title = {ACM Publication Policies and Plans},
year = {1959},
issue_date = {April 1959},
publisher = {Association for Computing Machinery},
address = {New York, NY, USA},
volume = {6},
number = {2},
issn = {0004-5411},
doi = {10.1145/320964.320965},
journal = {J. ACM},
month = apr,
pages = {121–122},
numpages = {2}
}

@article{Aaron1969,
author = {Finerman, Aaron},
title = {An Editorial Note},
year = {1969},
issue_date = {March 1969},
publisher = {Association for Computing Machinery},
address = {New York, NY, USA},
volume = {1},
number = {1},
issn = {0360-0300},
doi = {10.1145/356540.356541},
journal = {ACM Comput. Surv.},
month = mar,
pages = {1},
numpages = {1}
}

@article{Gotlieb1969,
author = {Gotlieb, C. C.},
title = {Editorial: On the ACM Publications},
year = {1969},
issue_date = {April 1969},
publisher = {Association for Computing Machinery},
address = {New York, NY, USA},
volume = {12},
number = {4},
issn = {0001-0782},
doi = {10.1145/362912.362916},
journal = {Commun. ACM},
month = apr,
pages = {197–198},
numpages = {2}
}

@article{Rice1975,
author = {Rice, John R.},
title = {Purpose and Scope},
year = {1975},
issue_date = {March 1975},
publisher = {Association for Computing Machinery},
address = {New York, NY, USA},
volume = {1},
number = {1},
issn = {0098-3500},
doi = {10.1145/355626.355627},
journal = {ACM Trans. Math. Softw.},
month = mar,
pages = {1–3},
numpages = {3}
}

@inproceedings{Akera2010,
author = {Akera, Atsushi and Mandelbaum, Mark},
title = {Mark Mandelbaum Interview: October 31 and December 29, 2009; New York City, NY},
year = {2010},
isbn = {9781450317719},
publisher = {Association for Computing Machinery},
address = {New York, NY, USA},
doi = {10.1145/1141880.1866271},
booktitle = {ACM Oral History Interviews},
numpages = {109}
}

@article{Hsiao1980,
author = {Hsiao, David K.},
title = {TODS---the First Three Years (1976--1978)},
year = {1980},
issue_date = {Dec.  1980},
publisher = {Association for Computing Machinery},
address = {New York, NY, USA},
volume = {5},
number = {4},
issn = {0362-5915},
doi = {10.1145/320610.320611},
journal = {ACM Trans. Database Syst.},
month = dec,
pages = {385–403},
numpages = {19}
}

@ARTICLE{Merger1962,

  author={},

  journal={Electrical Engineering}, 

  title={Agreement of merger: Approved hy the boards of directors of AIEE and IRE March 9, 1962}, 

  year={1962},

  volume={81},

  number={12},

  pages={4-7},

  doi={10.1109/EE.1962.6446673}}

@ARTICLE{Lee1996,

  author={J. A. N. {Lee}},

  journal={IEEE Annals of the History of Computing}, 

  title={"Those who forget the lessons of history are doomed to repeat it": or, Why I study the history of computing}, 

  year={1996},

  volume={18},

  number={2},

  pages={54-62},

  doi={10.1109/85.489724}}

@ARTICLE{BirthOfAConcept,

  author={},

  journal={IEEE Spectrum}, 

  title={Spectral lines: Birth of a concept}, 

  year={1971},

  volume={8},

  number={1},

  pages={21-22},

  doi={10.1109/MSPEC.1971.6501060}}

@ARTICLE{Huskey1968,

  author={{Huskey}, Harry D},

  journal={IEEE Transactions on Computers}, 

  title={Editor's notice [Transactions name change]}, 

  year={1968},

  volume={C-17},

  number={1},

  pages={1-1},

  doi={10.1109/TC.1968.5008861}}

@ARTICLE{ComputerPetition,

  author={},

  journal={Computer}, 

  title={On the Group's Petition for a Change to Society Status}, 

  year={1970},

  volume={3},

  number={5},

  pages={1-1},

  doi={10.1109/C-M.1970.216697}}

@INPROCEEDINGS{King2009,

  author={W. K. {King} and S. K. {Land}},

  booktitle={2009 IEEE Conference on the History of Technical Societies}, 

  title={A Historical Perspective of the IEEE Computer Society: Six Decades of Growth with the Technology It Represents}, 

  year={2009},

  volume={},

  number={},

  pages={1-6},

  doi={10.1109/HTS.2009.5337852}}

@ARTICLE{ComputerTheFirst25,

  author={Charles Concordia, Walter Anderson},

  journal={Computer}, 

  title={The First 25 Years}, 

  year={1976},

  volume={9},

  number={12},

  pages={41-53},

  doi={10.1109/C-M.1976.218469}}

@article{Thompson1977,
author = {Thompson, Howard K.},
title = {A SIGBIO File of Biomedical Computing Positions and Potential Candidates for Such Positions},
year = {1977},
issue_date = {March 1977},
publisher = {Association for Computing Machinery},
address = {New York, NY, USA},
volume = {2},
number = {1},
issn = {0163-5697},
doi = {10.1145/992264.992266},
journal = {SIGBIO Newsl.},
month = mar,
pages = {7–8},
numpages = {2}
}

@ARTICLE{FirstIEEETSE,

  author={},

  journal={IEEE Transactions on Software Engineering}, 

  title={Editor's notice}, 

  year={1975},

  volume={SE-1},

  number={1},

  pages={1-6},

  doi={10.1109/TSE.1975.6312815}}

@Book{naur1976software,
 author = {Naur, Peter},
 title = {Software engineering : concepts and techniques : proceedings of the NATO conferences},
 publisher = {Petrocelli/Charter},
 year = {1976},
 address = {New York},
 isbn = {978-0884053347}
 }

@article{Parnas1968,
author = {Parnas, David L.},
title = {Letters to the Editor: On Improving the Quality of Our Technical Meetings},
year = {1968},
issue_date = {Aug. 1968},
publisher = {Association for Computing Machinery},
address = {New York, NY, USA},
volume = {11},
number = {8},
issn = {0001-0782},
doi = {10.1145/363567.363568},
journal = {Commun. ACM},
month = aug,
pages = {537–538},
numpages = {2}
}

@misc{Yost2014,
author = {Jeffrey Yost and H. True Seaborn},
title = {TRUE SEABORN: An Interview Conducted by Jeffrey R. Yost, IEEE Computer Society, May 22, 2014},
year = {2014},
publisher = {IEEE Computer Society},
address = {Los Alamitos, CA},
url = {https://ethw.org/Oral-History:H._True_Seaborn},
}

@misc{SpringerHistory,
author = {Springer},
title = {Springer - Driving academic publishing since 1842},
year = {2017},
url = {https://www.springer.com/gp/about-springer/history}
}

@ARTICLE{Ritchie1984,  author={D. M. {Ritchie}},  journal={AT T Bell Laboratories Technical Journal},   title={The UNIX system: The evolution of the UNIX time-sharing system},   year={1984},  volume={63},  number={8},  pages={1577-1593},  doi={10.1002/j.1538-7305.1984.tb00054.x}}

@article{Denning1981,
author = {Denning, Peter J. and Feigenbaum, Edward and Gilmore, Paul and Hearn, Anthony and Ritchie, Robert W. and Traub, Joseph},
title = {A Discipline in Crisis},
year = {1981},
issue_date = {June 1981},
publisher = {Association for Computing Machinery},
address = {New York, NY, USA},
volume = {24},
number = {6},
issn = {0001-0782},
doi = {10.1145/358669.358682},
journal = {Commun. ACM},
month = jun,
pages = {370–374},
numpages = {5}
}

@misc{roberts2016,
author = {Eric Roberts},
Title = {A History of Capacity Challenges in Computer Science},
URL = {https://cs.stanford.edu/people/eroberts/CSCapacity.pdf},
year = {2016},
month = {03}
}

\end{document}